\documentclass[fleqn,usenatbib]{mnras}

\usepackage{newtxtext,newtxmath}

\usepackage[T1]{fontenc}

\DeclareRobustCommand{\VAN}[3]{#2}
\let\VANthebibliography\thebibliography
\def\thebibliography{\DeclareRobustCommand{\VAN}[3]{##3}\VANthebibliography}

\usepackage{graphicx}	
\usepackage{amsmath}	
\usepackage{lineno}     
\usepackage{color}

\title[Polarization Perspectives on Her X–1]{Polarization Perspectives on Hercules X–1: Further Constraining the Geometry}
\author[Q. C. Zhao et al.]{
Q. C. Zhao,$^{1,2}$
H. C. Li,$^{3}$\thanks{E-mail: hancheng.li@unige.ch}
L. Tao,$^{1}$\thanks{E-mail: taolian@ihep.ac.cn}
H. Feng,$^{4}$
S. N. Zhang,$^{1}$
R. Walter, $^{3}$
M. Y. Ge,$^{1}$
H. Tong,$^{5}$
L. Ji,$^{6}$
\newauthor{L. Zhang,$^{1}$
J. L. Qu,$^{1}$
Y. Huang,$^{1}$
X. Ma,$^{1}$
S. Zhang,$^{1}$
Q. Q. Yin,$^{1}$
H. X. Yin,$^{7}$
R. C. Ma,$^{1}$,
S. J. Zhao,$^{1,2}$}
\newauthor{P. P. Li,$^{1,2}$
Z. X. Yang,$^{8}$
H. X. Liu,$^{1}$
W. Yu,$^{1,2}$
Y. M. Huang,$^{1,2}$
Z. X. Li,$^{1,2}$
Y. J. Li,$^{1,2}$
J. Y. Xiao,$^{1,2}$
K. Zhao,$^{1,2}$
}
\\
$^{1}$Key Laboratory of Particle Astrophysics, Institute of High Energy Physics, Chinese Academy of Sciences, Beijing 100049, China\\
$^{2}$University of Chinese Academy of Sciences, Beijing 100049, China\\
$^{3}$Department of Astronomy, University of Geneva, 16 Chemin d’Ecogia, Versoix, CH-1290, Switzerland\\
$^{4}$Department of Astronomy, Tsinghua University, Beijing 100084, China\\
$^{5}$School of Physics and Materials Science, Guangzhou University, Guangzhou 510006, China\\
$^{6}$School of Physics and Astronomy, Sun Yat-sen University, Zhuhai, 519082, China\\
$^{7}$Shandong Key Laboratory of Optical Astronomy and Solar-Terrestrial Environment, School of Space Science and Physics, Institute of Space Sciences,\\
Shandong University, Weihai, Shandong 264209, China\\
$^{8}$School of Physics and Optoelectronic Engineering, Shandong Unversity of Technology, Zibo 255000, China\\
}

\date{Accepted 2024 April 27. Received 2024 March 29; in original form 2023 September 29}

\pubyear{2024}

\begin{document}
\label{firstpage}
\pagerange{\pageref{firstpage}--\pageref{lastpage}}
\maketitle

\begin{abstract}

We conduct a comprehensive analysis of the accreting X-ray pulsar, Hercules X-1, utilizing data from \textit{IXPE} and \textit{NuSTAR}. \textit{IXPE} performed five observations of Her X-1, consisting of three in the Main-on state and two in the Short-on state. Our time-resolved analysis uncovers the linear correlations between the flux and polarization degree as well as the pulse fraction and polarization degree. Geometry parameters are rigorously constrained by fitting the phase-resolved modulations of Cyclotron Resonance Scattering Feature and polarization angle with a simple dipole model and Rotating Vector Model respectively, yielding roughly consistent results. The changes of $\chi_{\rm p}$ (the position angle of the pulsar's spin axis on the plane of the sky) between different Main-on observations suggest the possible forced precession of the neutron star crust. Furthermore, a linear association between the energy of Cyclotron Resonance Scattering Feature and polarization angle implies the prevalence of a dominant dipole magnetic field, and their phase-resolved modulations likely arise from viewing angle effects.
\end{abstract}




\section{Introduction}
\label{sec:intro}
The accreting X-ray pulsar, Hercules X--1 (hereafter Her X--1), was initially discovered by the X-ray satellite \textit{Uhuru} in 1971 \citep{Tananbaum_Uhuru_HerX-1}. Her X--1 consists of a 1.5 $\rm M_{\odot}$ accreting neutron star and an A/F-type donor star, HZ Her, with a mass of 2.2 $\rm M_{\odot}$ \citep{Deeter_donor, Reynolds_donor}. The distance of Her X--1 has been estimated to be $\sim$7 kpc \citep{Bailer_distance}. As a luminous persistent source, Her X--1 has been regularly monitored and exhibits various intriguing variability features. The neutron star rotates with a period of $\sim$1.24 seconds and undergoes eclipses with a period of $\sim$1.7 days \citep{Tananbaum_Uhuru_HerX-1, Giacconi_spin}. The binary orbit is quasi-circular \citep{Staubert_eph}, with the orbital plane inclined at an angle estimated to be $>80^{\circ}$ \citep{Gerend_orbit_plane}. The system also demonstrates flux modulation with an approximate period of 35 days, which is also known as the super-orbital period \citep{Tananbaum_Uhuru_HerX-1, Giacconi_spin}. One cycle of such a period comprises a Main-on state and a Short-on state lasting for approximately 10 days and 5 days, respectively, both of which are followed by a distinct Off state of $\sim$10 days \citep{Scott_35days}. 
The nature of the super-orbital period remains a topic of ongoing debate. One plausible explanation is the precession of the warped accretion disk \citep{Giacconi_spin, Scott_35days, Ramsay_warpdisk, Zane_warpdisk}. Another hypothesis suggests that the precession of the neutron star may be responsible for this phenomenon \citep{Brecher_neutron_precession, Postnov_precession,Kolesnikov_neutron_precession}. Moreover, \citet{Staubert_profiles} discovered an additional synchronization cycle with a periodicity of approximately 35 days through a comprehensive investigation of the $\sim$ 1.24 s pulse profiles. This synchronization cycle was attributed to the precession of the neutron star. 

The X-ray spectra of Her X-1 follow a power-law continuum with an exponential cutoff, which is a characteristic feature commonly observed in accreting binary pulsars \citep{BW_model, Staubert_crsf}. Her X--1 exhibits a broad Cyclotron Resonance Scattering Feature (CRSF) in its hard X-ray spectrum. This feature arises from the resonant scattering of photons by electrons at Landau levels in a strong magnetic field ($\geq 10^{12}$ Gauss). The presence of a cyclotron line at $\sim$37 keV was initially detected through a balloon observation in 1976 \citep{Truemper_crsf}. Subsequently, the CRSF of Her X-1 has been extensively studied using various missions, including \textit{RXTE} \citep{Staubert_correlation_crsf, Vasco_correlation_crsf, Vasco_spectra}, \textit{INTEGRAL} \citep{Klochkov_integral}, \textit{Suzaku} \citep{Enoto_suzaku}, \textit{NuSTAR} \citep{Furst_nustar}, \textit{AstroSat} \citep{Bala_crsf}, and \textit{Insight}-HXMT \citep{Xiao_crsf}. \citet{Staubert_correlation_crsf} reported a positive linear correlation between the cyclotron line energy and the maximum X-ray flux observed during the corresponding Main-On state. This correlation was subsequently reaffirmed by \citet{Vasco_correlation_crsf}. Based on the established correlation, a continuous decay in the cyclotron line energy between the years 1996 and 2012 has been observed. However, since that period, the cyclotron line energy appears to follow a stable trend in its long-term evolution \citep{Staubert_crsf_long_term, Ji_crsf, Xiao_crsf, Staubert_crsf, Bala_crsf}. Besides, the cyclotron line energy is strongly dependent on the rotation phase of the neutron star \citep{Vasco_spectra, Furst_nustar}. \citet{Vasco_spectra} suggested further an apparent non-dependence of variation of cyclotron line energy with pulse phase on the 35-day phase, which could provide a way to constrain the geometric information of the system. The variation of cyclotron line energy with pulse phase may be attributed to the changing viewing angle from which the X-ray emitting regions are observed \citep{Kreykenbohm_pulse_phase_crsf}. For example, \citet{Suchy_dipole_crsf} have employed the simple dipole model and successfully reproduced the variation of the cyclotron line energy as a function of the pulse phase in GX 301--2.

X-ray polarization provides a new probe to diagnose theoretical models. For accreting X-ray pulsars (XRPs), it is predicted that the emitted radiation would exhibit high polarization degrees (PD) up to 60\%-80\% \citep{Caiazzo_polarization_model}. However, recent observations conducted by Imaging X-ray Polarimetry Explorer (\textit{IXPE}) for several XRPs, including Her X-1 \citep{Victor_Herx-1, Garg_herx-1,Heyl_herX-1}, Cen X-3 \citep{Tsygankov_cenX-3}, GRO J1008-57 \citep{Tsygankov_groj1008}, 4U 1626--67 \citep{Marshall_4U1627}, X Persei \citep{Mushtukov_xpeisei}, Vela X--1 \citep{Forsblom_velaX-1}, EXO 2030+375 \citep{Malacaria_exo2030}, GX 301-2 \citep{Suleimanov_gx301}, and LS V +44 17 \citep{Victor_LSV44}, have revealed unexpectedly low PDs. According to \citet{Victor_Herx-1}, the observed low PDs in Her X-1 could be the result of a combination of several mechanisms. These mechanisms include the inverse temperature profile of the neutron star atmosphere, the propagation of initially polarized X-rays through the magnetosphere of the neutron star, and the combination of emissions from the two magnetic poles. \citet{Long_0535} reported a non-detection polarization in 3--8 keV with Polarlight for 1A 0535+262. Notably, during this observation, the source was in a super-critical state. They propose that the relatively low PDs observed in accreting pulsars may not solely stem from the source being in a super-critical/sub-critical state. Instead, they suggest that these low PDs might represent a general feature of accreting pulsars.

In the case of Her X-1, \citet{Victor_Herx-1} conducted a comprehensive analysis using data from the first \textit{IXPE} observation. Furthermore, they effectively constrained the pulsar's geometry parameters by applying the Rotating Vector Model (RVM). More recently, \citet{Heyl_herX-1} attempted to unveil the precession of the neutron star crust by employing the RVM with data from the first three \textit{IXPE} observations. Additionally, \citet{Garg_herx-1} carried out a flux-resolved polarimetric analysis using the initial three \textit{IXPE} observations. They observed a significantly higher PD in the Short-on state compared to the Main-on state, potentially caused by the obstruction of the warped disk. In this paper, we conduct a comprehensive analysis of five \textit{IXPE} observations of Her X--1 to investigate its polarization properties. Additionally, we attempt to constrain the geometry parameters in two different ways: polarization and cyclotron line energy. Furthermore, we explore the correlations between the polarization angle and the cyclotron line energy, which are indicative of the magnetic field properties. The paper is organized as follows. We describe the observations information and data reduction methods in Section \ref{sec:data}, present results in Section \ref{sec:results}, and discuss the results in Section \ref{sec:discussion}.

\begin{table*}
\begin{tabular}{|l|l|l|l|l|l|l|}
\hline
\textbf{Instrument} & \textbf{ObsID} & \textbf{Tstart} & \textbf{$\phi_{35}$} & \textbf{Exposure (ks)} & \textbf{Turn on (MJD)} & \textbf{Spin period (s)}\\ \hline
IXPE                & 01001899      & 2022-02-17       & -0.03-0.18            & 255.9                   & 59628.50               &  1.2377094(1)  \\ 
                    & 02003801      & 2023-01-18       & 0.67-0.74            & 148.3                  & 59939.54               &  1.2377041(9)\\ 
                    & 02004001      & 2023-02-03       & 0.13-0.27            & 244.6                  & 59974.39               &  1.2377041(9)\\ 
    & 02003901 & 2023-07-09 &  0.60-0.69 & 145.4   &  60113.79 & 1.2377030(3)   \\
    & 02004101 & 2023-07-25 &  0.05-0.20  & 236.5   &  60148.64 & 1.2377030(3)   \\ \hline
NuSTAR               & 30402034006  & 2019-02-18       & 0.76-0.78            & 22.9                   & 58515.64               &  1.2377201(2)\\ 
    & 30601012002  & 2021-05-17       & 0.07-0.09            & 24.3                   & 59348.71               &  1.2377120(9)\\ 
                    & 90701330002  & 2021-10-06       & 0.17-0.19            & 24.0                   & 59488.04               &  1.2377110(1)\\ \hline
\end{tabular}
\caption{Basic information of the observations.}
\label{table:table1}
\end{table*}

\begin{table*}
\begin{tabular}{|l|l|l|l|l|l|l|l|l|}
\hline
\textbf{Method} & \textbf{Observations} & \textbf{1001899} & \textbf{2003801} & \textbf{2004001} & \textbf{02003901}  & \textbf{02004101}\\ \hline
                & $\phi_{35}$ & -0.03-0.18 & 0.67-0.74 & 0.13-0.27 & 0.60-0.69 & 0.05-0.20         \\ \hline
Pcube      & PD (\%) & 8.63$\pm$0.62 & 18.2$\pm$2.3 & 8.61$\pm$0.60  & 19.2$\pm$1.4    & 10.08$\pm$0.61          \\ 
                & PA ($^{\circ}$) & 59.2$\pm$2.1 & 42.5$\pm$3.7 & 48.0$\pm$2.0 & 41.8$\pm$2.0  & 52.6$\pm$1.7        \\ \hline
Xspec           & PD (\%) & 8.67$\pm$0.80 & 18.4$\pm$3.0 & 8.10$\pm$0.76 & 18.0$\pm$1.7  & 10.00$\pm$0.78          \\ 
                & PA ($^{\circ}$) & 60.3$\pm$2.6  & 41.6$\pm$4.7 & 47.9$\pm$2.7 & 41.9$\pm$2.8 & 51.6$\pm$2.2            \\ \hline

\end{tabular}
\caption{PD and PA of all DUs averaged estimated by the \texttt{PCUBE} algorithm and \texttt{XSPEC} for all five observations.}
\label{table:table2}
\end{table*}

\begin{figure*}
    \includegraphics[width=0.95\textwidth]{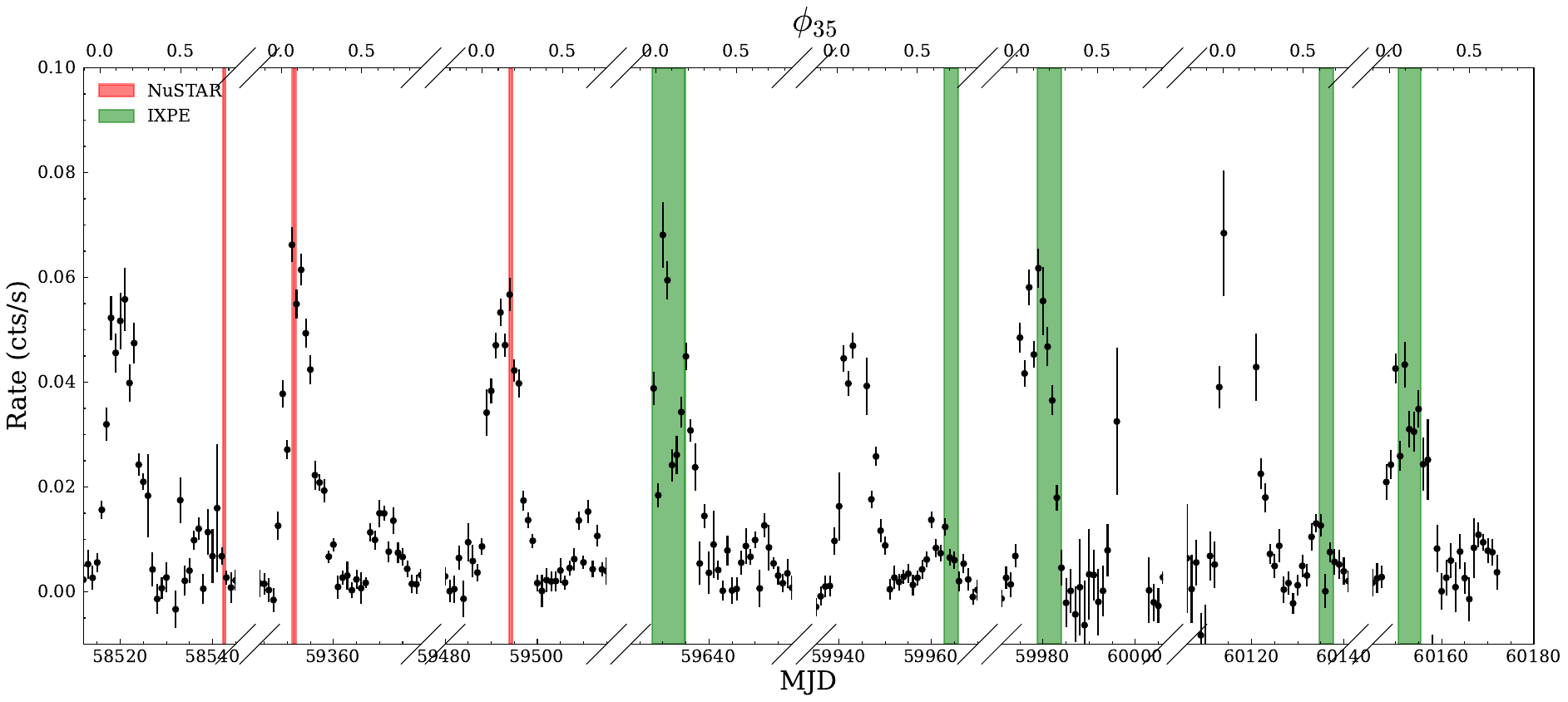}
    \caption{The \textit{Swift}-BAT daily lightcurve. The green strip and red strip indicate the time windows of \textit{IXPE} and \textit{NuSTAR} observations, respectively. Note that the effective exposures to Her X--1 by IXPE are discontinuous among these time windows.}
    \label{fig:fig1}
\end{figure*}

\begin{figure}
    \includegraphics[width=0.5\textwidth]{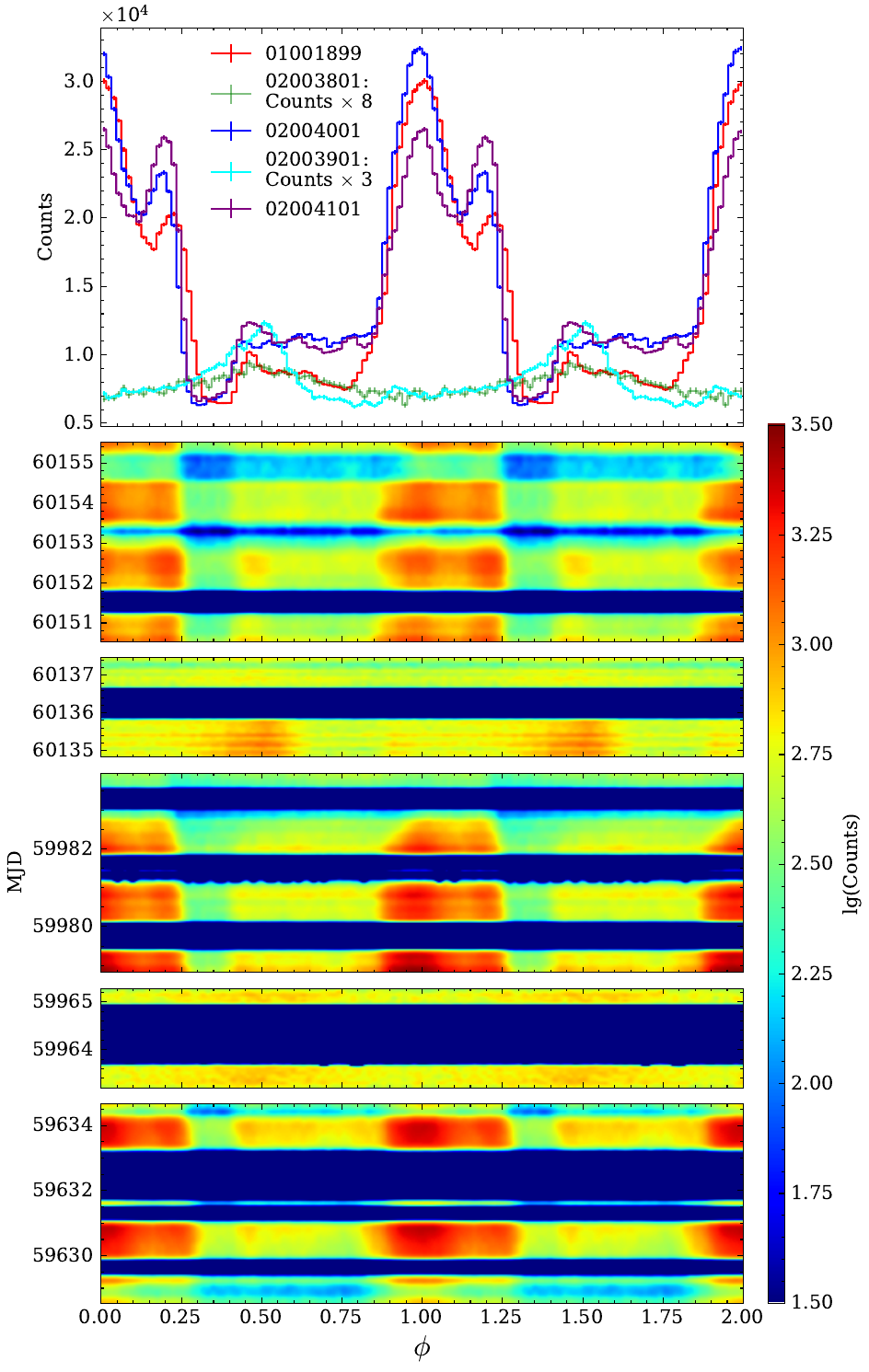}
    \caption{Top panel: The profiles of the three \textit{IXPE} observations over 2--8 keV (per 1/64 phase interval). Bottom panel: The two-dimensional maps illustrate the evolution of the pulse profiles with time for all \textit{IXPE} observations.}
    \label{fig:fig2}
\end{figure}

\begin{figure}
    \includegraphics[width=0.5\textwidth]{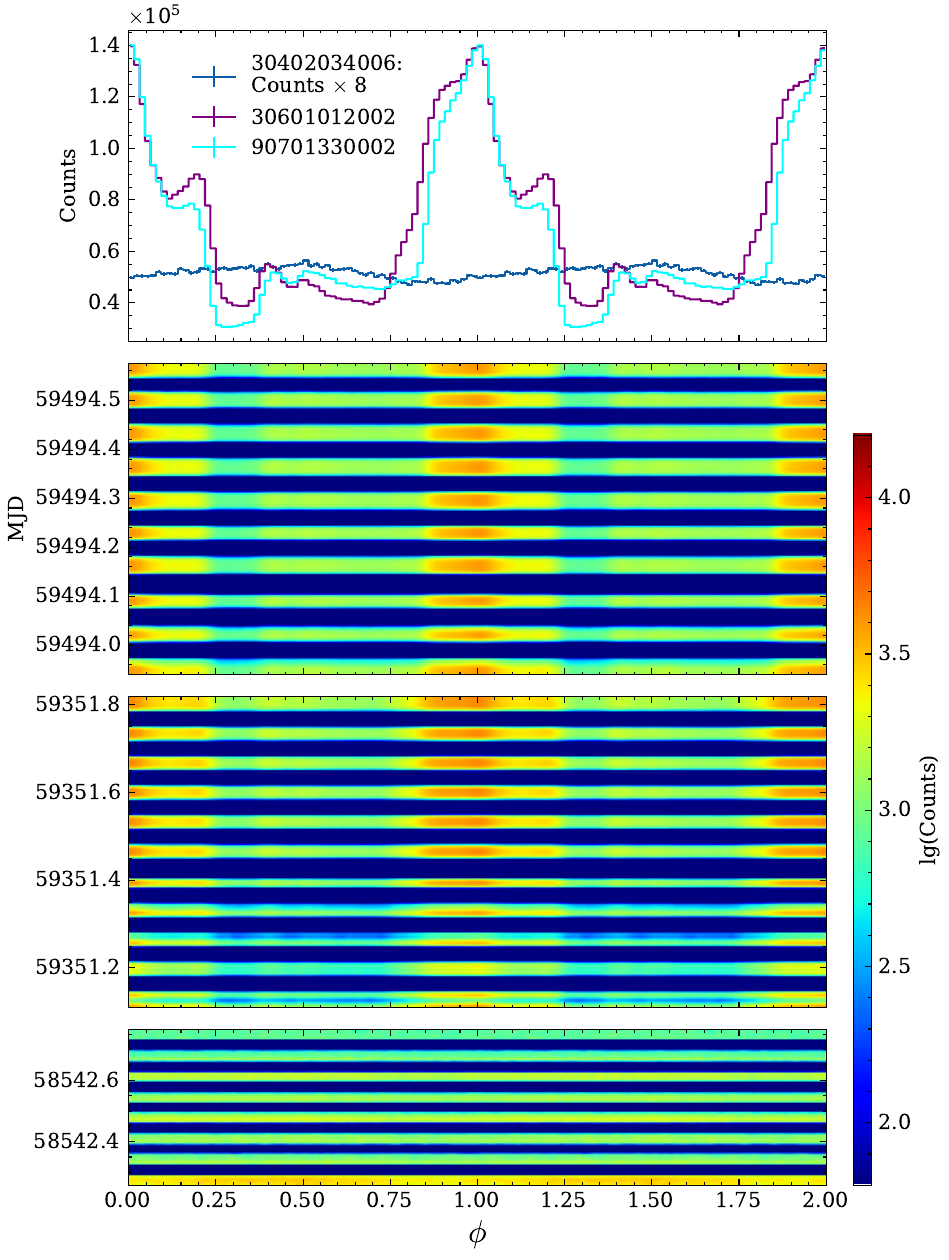}
    \caption{Same with Figure~\ref{fig:fig2}, but with \textit{NuSTAR} data (3--79 keV).}
    \label{fig:fig3}
\end{figure}

\begin{figure}
    \includegraphics[width=0.45\textwidth]{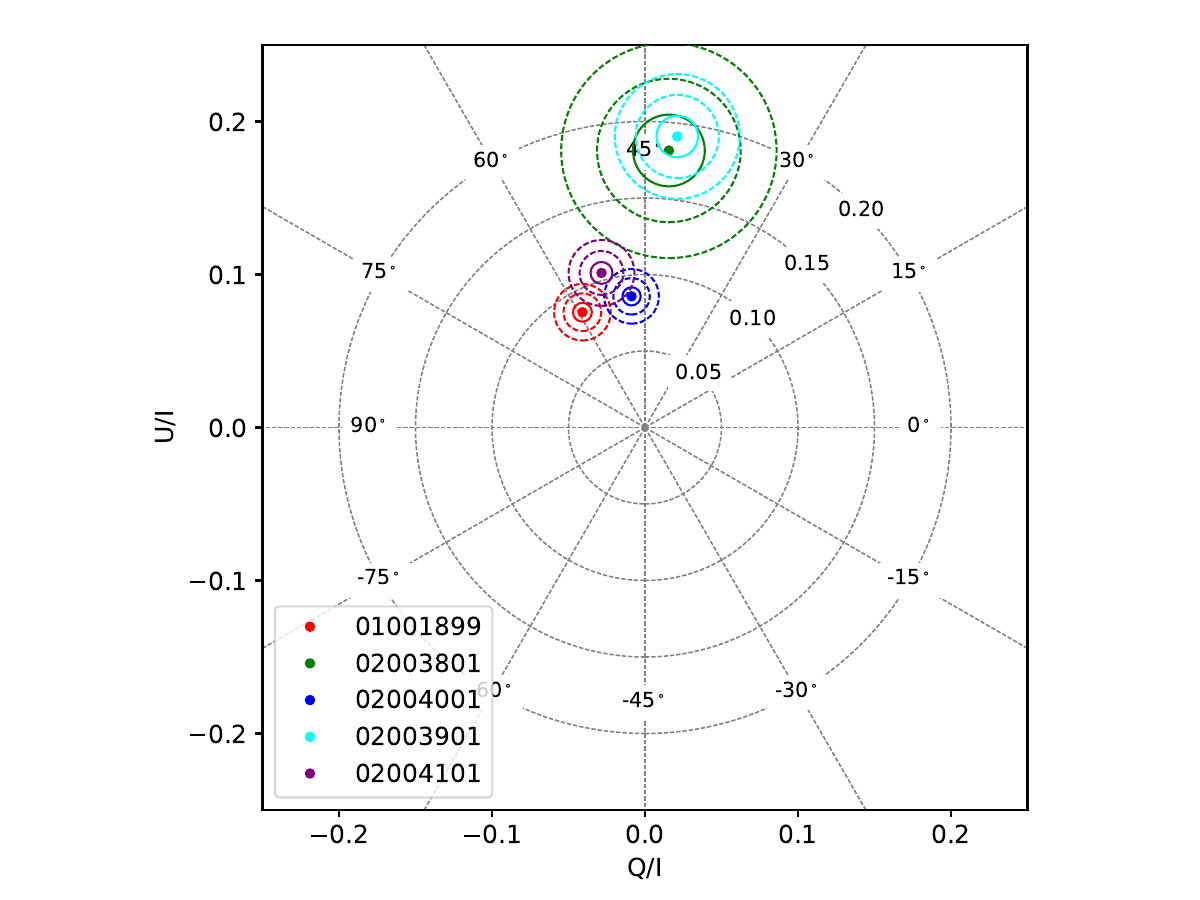}
    \caption{Pulse phase-averaged normalized Stokes paremeters U/I and Q/I over the energy range of 2-8 keV. The red, green, blue, cyan and purple colors correspond to the ObsIDs 01001899, 02003801, 02004001, 02003901, and 02004101. The contours in the plot depict the confidence levels at 1$\sigma$, 2$\sigma$, and 3$\sigma$.}
    \label{fig:fig4}
\end{figure}

\begin{figure*}
    \centering
    \includegraphics[width=0.95\textwidth]{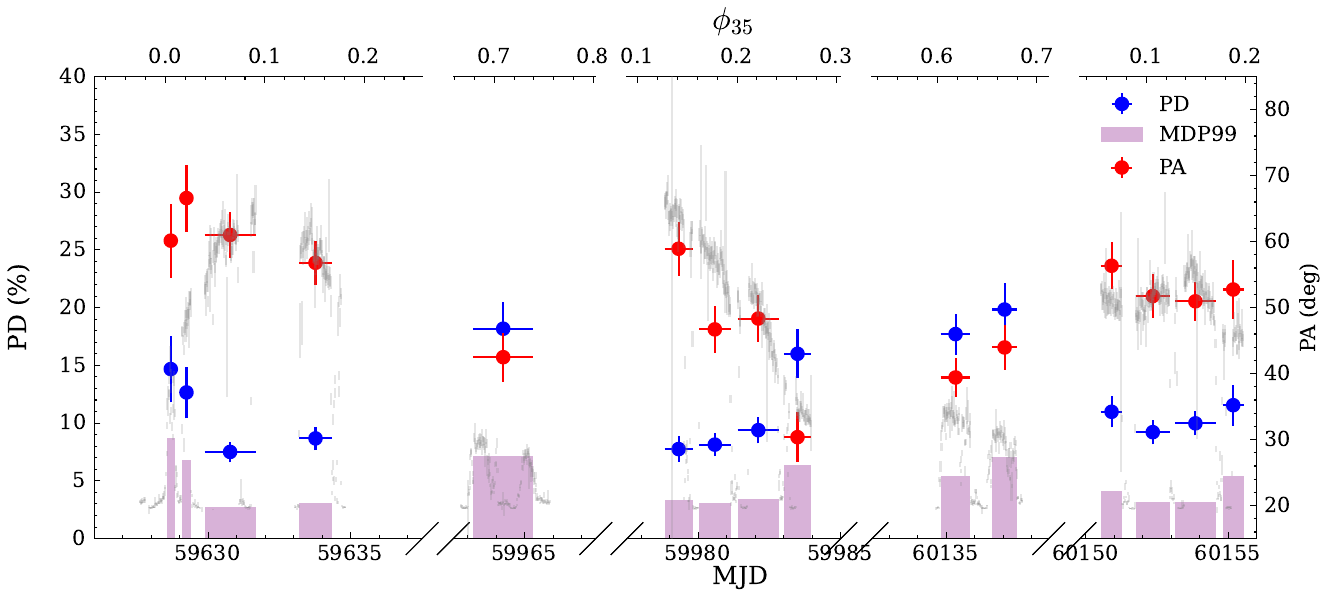}
    \caption{The PD, PA, and flux as a function of time /$\phi_{35}$ for all three observations of \textit{IXPE} are shown. The gray color represents the \textit{IXPE} rates with a time bin of 1000\,s. The blue and red color points represent the PD and PA, respectively. The purple strips indicate the minimum detectable polarization at the 99\% confidence level (MDP99). Eclipses and pre-eclipse dips are excluded from the analysis.}
    \label{fig:fig5}
\end{figure*}

\begin{figure*}
    \centering
    \includegraphics[width=0.95\textwidth]{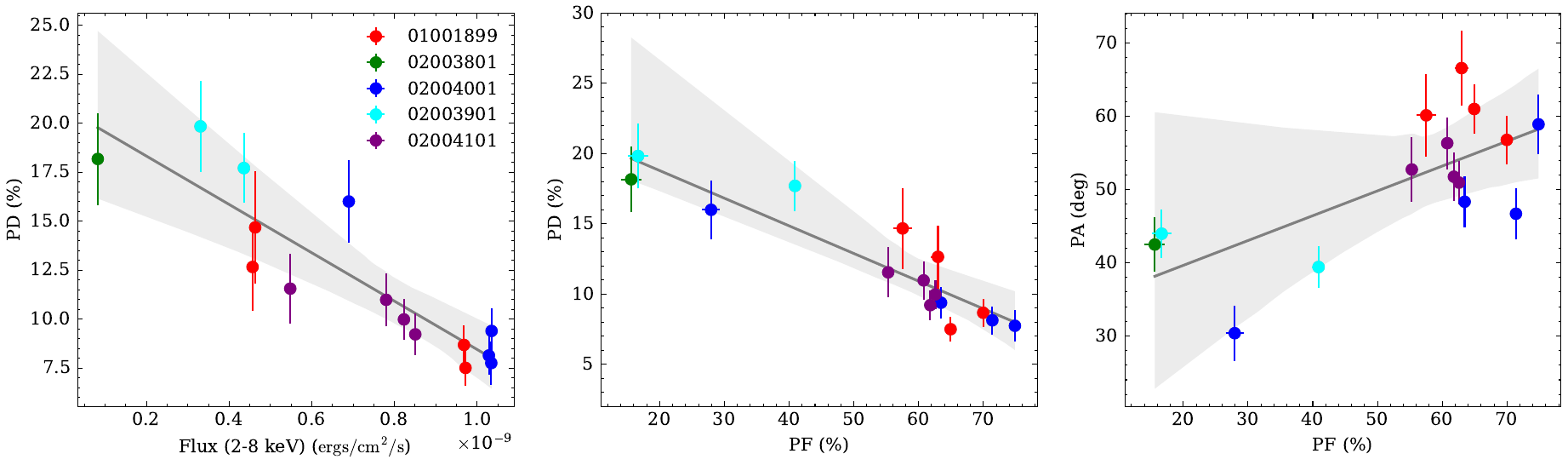}
    \caption{Left Panel: the PD as a function of Flux. The gray line illustrates the linear regression of PD against PF, while the shaded area indicates the 3$\sigma$ confidence interval of the fittings. The Pearson correlation coefficient is -0.89. Middle panel: the PD as a function of PF. The slope and intercept are -0.20 and 22.7, respectively. The Pearson correlation coefficient is -0.92. Right panel: the PA as a function of PF. The purple line illustrates the linear regression of the PA against PF, while the shaded purple region corresponds to the 3$\sigma$ confidence interval of the fittings. The slope and intercept are 0.33 and 33.3. The Pearson correlation coefficient is 0.69. The $p$-values with the null hypothesis that the slope equals to zero are $8 \times 10^{-6}$, $2 \times 10^{-6}$ and 0.005 for the left, middle and right panels, respectively. PD and PA are estimated by \texttt{PCUBE}}.
    \label{fig:fig6}
\end{figure*}

\section{OBSERVATIONS AND DATA REDUCTION}
\label{sec:data} 

\subsection{IXPE}
\label{sec:data:ixpe}

The Imaging X-ray Polarimetry Explorer (\textit{IXPE}) has so far performed five observations of Her X--1, as shown in Figure~\ref{fig:fig1}. Three observations are in the Main-on state (01001899, 02004001, 02004101) while the other two are in the Short-on state (02003801, 02003901). The observational information is listed in Table \ref{table:table1}. We start our analysis from level-2 data products which are downloaded from \textit{HEASARC ARCHIVE WEBSITE} \footnote{\url{https://heasarc.gsfc.nasa.gov/cgi-bin/W3Browse/w3browse.pl}}. 

For the first observation: ObsID 01001899, the position offsets and energy calibration offsets need to be corrected as outlined in the \texttt{readme} file. The position offsets are corrected using the \texttt{fmodhead} tool within \texttt{ftools}, while the energy calibration offsets are corrected using the \texttt{xppicorr} tool provided in \texttt{ixpeobssim} \citep{Baldini_obssim}. We apply the Solar System barycentric correction to the photon arrival times with the \texttt{barycorr} tool within \texttt{ftools} for all five observations. The binary orbit correction is corrected by using the ephemeris reported by \citet{Staubert_eph}. A circular region with a radius of 96 arcsec is selected as the source region. We do not subtract the background since it has no significant effects on Her X--1, which is a relatively bright source \citep{DiMarco_background}.

We select events from the source region within the energy range of 2--8 keV. The polarimetric analysis is conducted using the \texttt{ixpeobssim} software, which allows for a model-independent analysis. The \texttt{PCUBE} algorithm is employed to generate the polarization cubes. Additionally, we perform a spectropolarimetric analysis using the \texttt{PHA1}, \texttt{PHA1Q}, and \texttt{PHA1U} algorithms to generate the Stokes spectra. These spectra are then fitted in \texttt{XSPEC} \citep{Arnaud_xspec} to obtain the polarization degrees (PD) and polarization angle (PA). Eclipses and pre-eclipse dips are excluded from the analysis.

\subsection{NuSTAR}
\label{sec:data:nustar}

The Nuclear Spectroscopic Telescope Array (\textit{NuSTAR}) \citep{Harrison_nustar} has carried out many observations on Her X--1 to reveal the long-term evolution of cyclotron line \citep{ Staubert_crsf}. In the absence of simultaneous observations between \textit{NuSTAR} and \textit{IXPE}, the two Main-on \textit{NuSTAR} observations of Her X--1 (ObsIDs 30601012002 and 90701330002) and one Short-on observation (ObsID 30402034006) from the archived data, are selected for analysis. The turn-on time is estimated from \textit{Swift}-BAT orbital lightcurve following the method described by \citet{Staubert_35day}.

The cleaned level-2 events products are extracted using the \texttt{nupipeline} routine of the \textit{NuSTAR} Data Analysis Software (\texttt{NuSTARDAS}) package, which is distributed within the \texttt{HEASOFT 6.31.1} software. Additionally, barycentric correction and binary orbit correction are applied. The source events are extracted from a circular region with a radius of 80 arcsec, while the background events are extracted from an annulus region with an inner radius of 90 arcsec and an outer radius of 120 arcsec. We produce the spectra by using the \texttt{nuproducts} to make further analysis. The spectra are rebinned with a minimum of 25 counts per bin, and the used energy band is 3--79 keV.

\section{ANALYSIS AND RESULTS}
\label{sec:results}

\subsection{Timing analysis}

The frequencies are determined using the epoch-folding technique \citep{Leahy_epoch_folding}. The estimated spin periods are listed in Table~\ref{table:table1}. These periods are then utilized to calculate the pulse phases, which are subsequently used to fold the profiles. In Figure~\ref{fig:fig2}, we present the profiles of all \textit{IXPE} observations and the corresponding evolution with time. 
Noticeably, the profiles exhibit variations between the Main-on state and Short-on state, suggesting a potential influence from the obscuration of the precessing accretion disk. Furthermore, even within the Main-on state, discernible profile differences can be observed, potentially signifying the influence of neutron star precession \citep{Staubert_profiles}. In the case of ObsID 02004101, the peak at phase $\sim$ 0.2/1.2 exhibits a greater intensity compared to the other two Main-on state observations. This disparity could arise from alterations in the relative flux contributions of different beam patterns (fan or pencil), or potentially due to changes in the system's geometry, like the precession of the neutron star. The profiles from the two \textit{NuSTAR} observations are presented in Figure~\ref{fig:fig3}. The profiles of the Main-on state are more consistent with the profiles from the initial two \textit{IXPE} Main-on observations.

\subsection{Polarimetric analysis}
\label{sec:results:time_resolved}

\subsubsection{Time resolved polarimetric analysis}

In Figure~\ref{fig:fig4}, we present the polarization parameters for all \textit{IXPE} observations: PD (polarization degree) and PA (polarization angle). These parameters are estimated using the model-independent \texttt{PCUBE} algorithm. Significant polarization is detected in all five \textit{IXPE} observations. The PD/PA are 8.63$\pm$0.62\%/59.2$\pm$2.1$^\circ$, 18.2$\pm$2.3\%/42.5$\pm$3.7$^\circ$, 8.61$\pm$0.60\%/48.0$\pm$2.0$^\circ$, 19.2$\pm$1.4/41.8$\pm$2.0$^\circ$ and 10.08$\pm$0.61/52.6$\pm$1.7$^\circ$ for ObsID 01001899, 02003801, 02004001, 02003901 and 02004101, respectively. In a general trend, the PD during the Short-on state is notably higher compared to those during the Main-on state. We also use the spectro-polarimetric method to estimate the polarization parameters. We fit the Stokes spectra: I, Q, U with the model: \texttt{Const*Polconst*Powerlaw} in \texttt{XSPEC}. The results are generally consistent with the model-independent \texttt{PCUBE} algorithm, as demonstrated in Table \ref{table:table2}.

We divide the data into multiple segments to investigate the evolution of PD and PA with time or $\phi_{35}$. The PD of Short-on state observations is higher than the PD of Main-on state observations as shown in Figure~\ref{fig:fig5}. In ObsID 01001899, the PD of the first two segments which is at the beginning of the Main-on is higher than other segments. Besides, the PD is also higher at the end of Main-on (the 4th segment) as shown in the ObsID 02004001. Our results are in agreement with the previous studies in \citet{Victor_Herx-1} and \citet{Garg_herx-1}, taking into account the uncertainties. We also identify a linear anti-correlation between the flux and the PD as shown in Figure~\ref{fig:fig6}. The Pearson correlation coefficient is -0.89.

We also calculate the pulsed fraction (PF) to examine the correlation between the polarization parameters and PF. PF is defined as: 
\begin{equation}
\mathrm{PF} = \frac{{F_{\rm max}} - {F_{\rm min}}}{{F_{\rm max}} + {F_{\rm min}}}.
\end{equation}

As depicted in Figure~\ref{fig:fig6}, the observations of Short-on state: ObsIDs 02003801 and 02003901 exhibit significantly lower PF compared to the Main-on state observations. The observations of the Main-on state demonstrate higher and comparable PF. An anti-correlation is observed between PD and PF. We carry out a linear regression, which yields a slope of -0.20 and an intercept of 22.7. The Pearson correlation coefficient is -0.92. Moreover, we apply a linear regression to examine the correlation between PA and PF. The resulting slope is 0.33, and the intercept is 33.3. The Pearson correlation coefficient is 0.69. 

\subsubsection{Phase resolved polarimetric analysis}
\label{sec:results:phase_polarization}
\begin{figure*}
    \centering
    \includegraphics[width=0.85\textwidth]{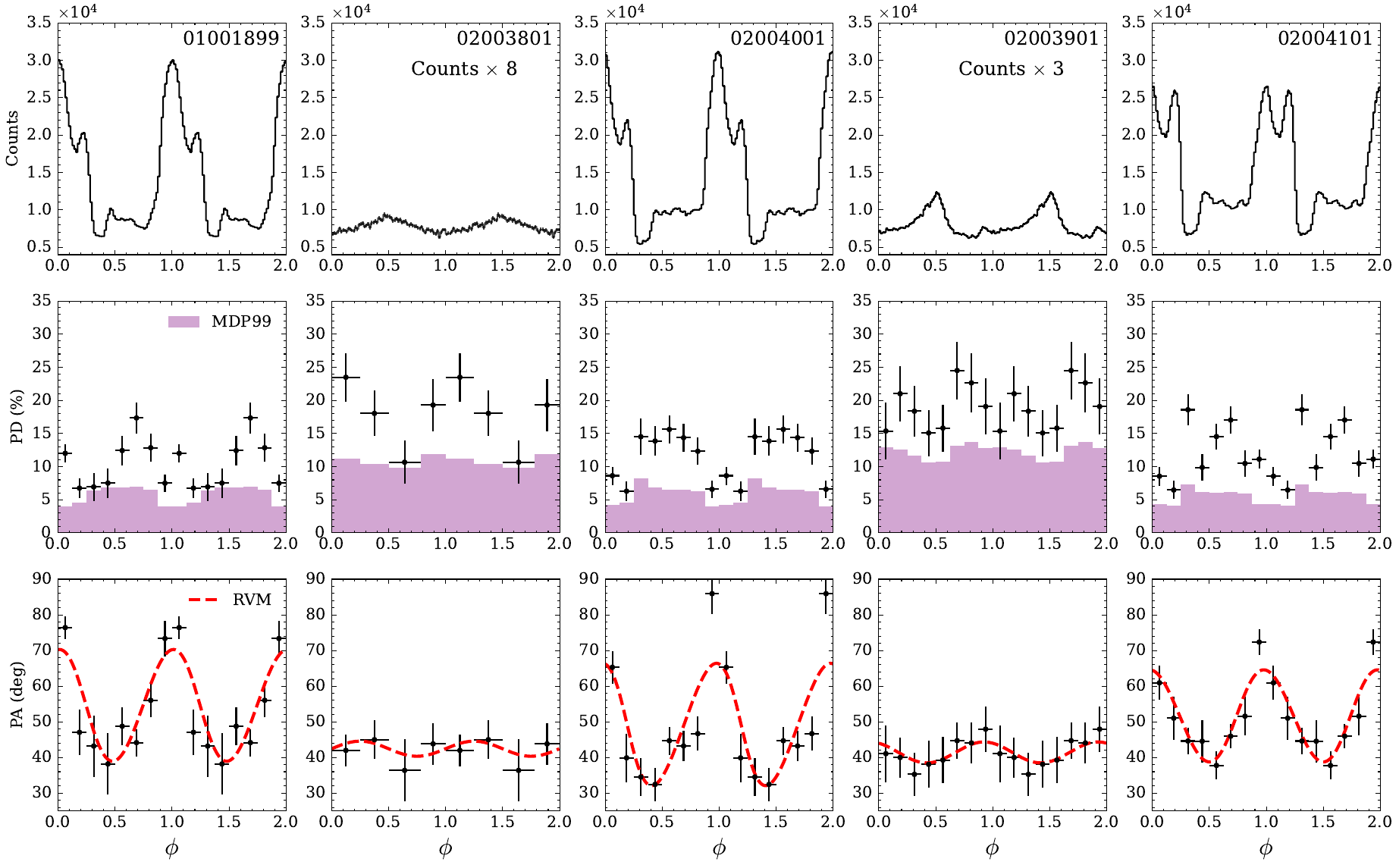}
    \caption{Top panel: Profiles of three \textit{IXPE} observations. Middle panel: Variation of PD with pulse phase. Bottom panel: Variation of PA with pulse phase. The red color curves represent the RVM.}
    \label{fig:fig7}
\end{figure*}

We divide the pulse phase into multiple bins to study the variations of PD and PA with respect to the pulse phase. From the time-resolved results of the obsID 02004001 (Figure \ref{fig:fig5}), we notice a distinctly higher PD on the 4th segment, which is at $\phi_{35}\sim0.3$ where the disk approximately cuts across the neutron star face again \citep{Scott_35days}; so we exclude such time segment in obsID 02004001 to avoid contamination in phase-resolved analysis. The PD of each bin is higher than the minimum detectable polarization at the 99\% confidence level (MDP99) to make sure the measurements of PA are reliable. As depicted in Figure~\ref{fig:fig7}, the PD values for the Main-on state observations range from approximately 7\% to 19\%, while for the Short-on observations, they range from approximately 11\% to 26\%. Additionally, the PA variations with pulse phase display distinct modulations, particularly in the case of the three observations during the Main-on state. While for the Short-on state observations, the amplitudes of PA modulations with pulse phase appear relatively small compared to those observed during the Main-on state. The PD exhibits an anti-correlation with the pulse phase: when the flux reaches its peak, the PD approaches its minimum value.

\subsection{Spectral analysis}
\label{sec:results:spec}
\subsubsection{Phase averaged spectroscopy}

\begin{figure}
    \centering
    \includegraphics[width=0.45\textwidth]{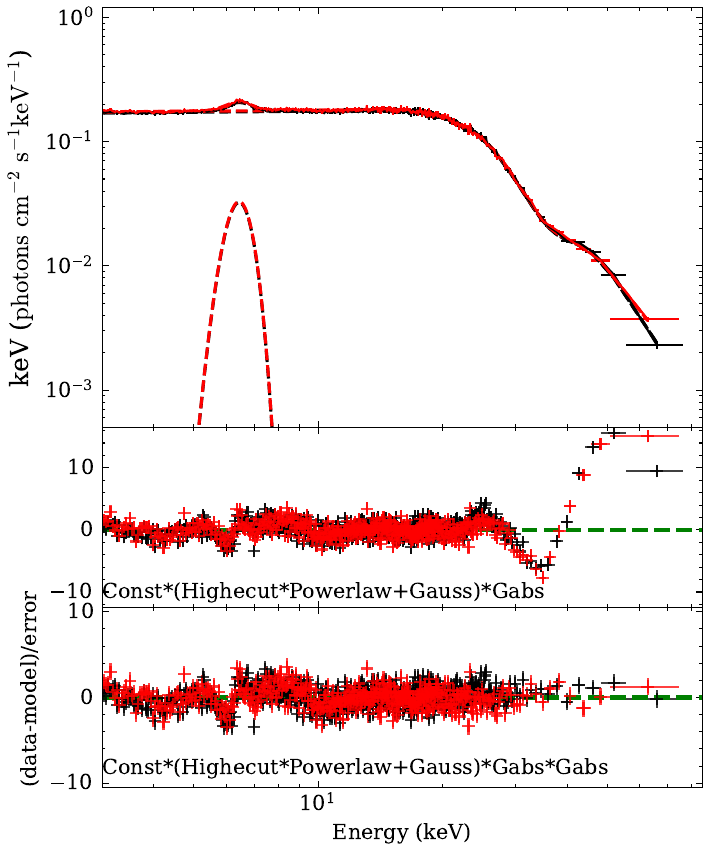}
    \caption{Representative \textit{NuSTAR} spectra of ObsID 30601012002. The black color corresponds to the FPMA data, while the red color corresponds to the FPMB data. The middle and bottom panels show the spectral fitting residuals with \texttt{Const*(Highecut*Powerlaw+Gauss)*Gabs} and \texttt{Const*(Highecut*Powerlaw+Gauss)*Gabs*Gabs}, respectively. The CRSF can be clearly seen in the middle panel. We add another \texttt{Gabs} model to fit the CRSF.}
    \label{fig:fig8}
\end{figure}

In the spectral analysis of \textit{NuSTAR} observations, we employ the well-established phenomenological model \texttt{Highecut*Powerlaw}, which has proven effective in previous investigations of Her X-1's spectral properties. The \texttt{Highecut} model is recognized for its distinct feature—an abrupt dip at the cutoff energy, as extensively discussed in previous studies \citep{Kretschmar_cutoff, Kreykenbohm_cutoff, Coburn_cutoff}. To account for this characteristic, we incorporate a \texttt{Gabs} model into our analysis. Additionally, we utilize a \texttt{Gauss} model to fit the iron fluorescence line, with the centroid line energy fixed at 6.4 keV. In Figure~\ref{fig:fig8}, we observe a clear cyclotron resonance scattering feature within the $\sim$30–40 keV range, which we address by introducing another \texttt{Gabs} model. To adjust calibration discrepancies between FPMA and FPMB (with FPMA fixed at unity as a reference), we incorporate a \texttt{Const} model. In summary, our final model expression is given as: \texttt{Const*(Highecut*Powerlaw+Gauss)*Gabs*Gabs}.

\begin{figure}
    \centering
    \includegraphics[width=0.45\textwidth]{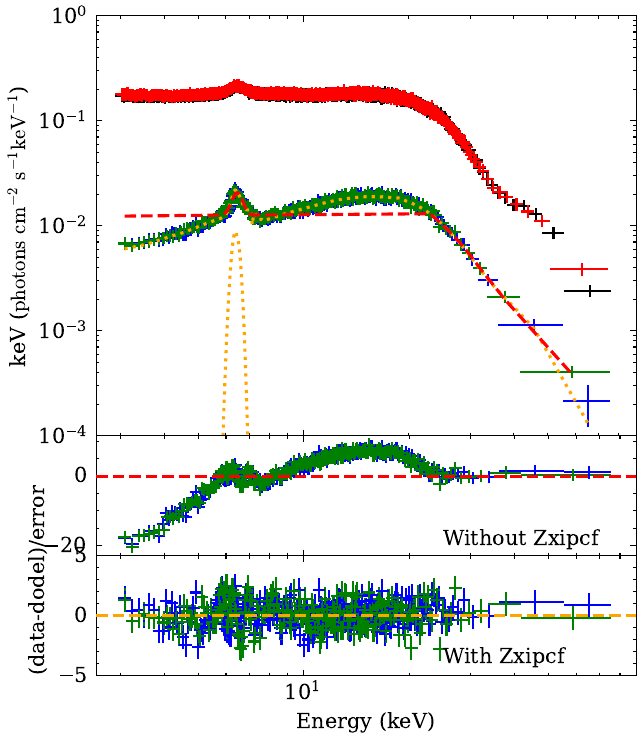}
    \caption{Top panel: the spectra of Main-on state and Short-on state. The black and red color represent the FPMA and FPMB spectra from the ObsID 30601012002, which is in the Main-on state. The blue and green represent the Short-on spectra (ObsID 30402034006) for FPMA and FPMB, respectively. The red dashed and yellow dotted lines represent the models with and without the \texttt{Zxipcf} component respectively, and the fitting residuals of them are shown in the middle panel and bottom panel respectively.} 
    \label{fig:fig9}
\end{figure}

A notable discrepancy in spectral hardness between the Short-on and Main-on states is evident in Figure~\ref{fig:fig9}. Such discrepancy is widely attributed to the 35-day flux modulation, a consequence of the obscuration by the precessing warped disk, thereby inducing distinctive spectral variations in these states. Our initial attempt involved fitting the Short-on state spectrum with fixed photon indices derived from the Main-on state. However, this approach yielded an unsatisfactory fit. Subsequently, we refined the fitting procedure by introducing a \texttt{Zxipcf} component, which represents a partial covering absorption model. Incorporating this component resulted in an improved fit, suggesting that the 35-day flux modulations likely arise from the absorption effects associated with the presence of the disk wind.

\subsubsection{Phase resolved spectroscopy}

\begin{figure}
    \centering
    \includegraphics[width=0.40\textwidth]{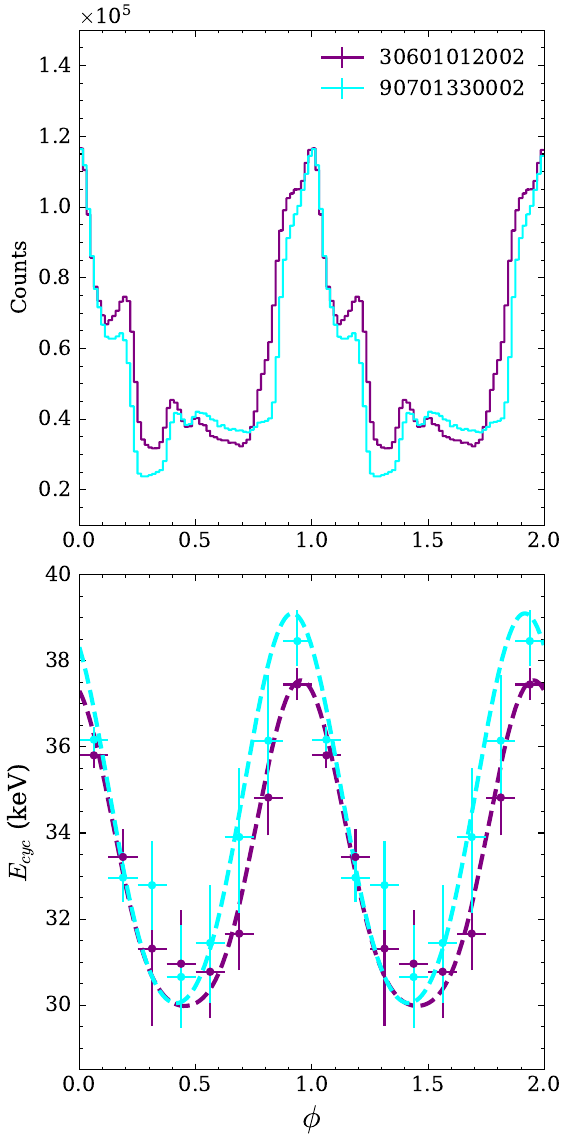}
    \caption{Top panel: Profiles of the \textit{NuSTAR} observations. Bottom panel: Variation of the cyclotron line energy as a function of the pulse phase. The dashed curves represent the simple dipole model.}
    \label{fig:fig10}
\end{figure}

\begin{figure*}
    \centering
    \includegraphics[width=0.85\textwidth]{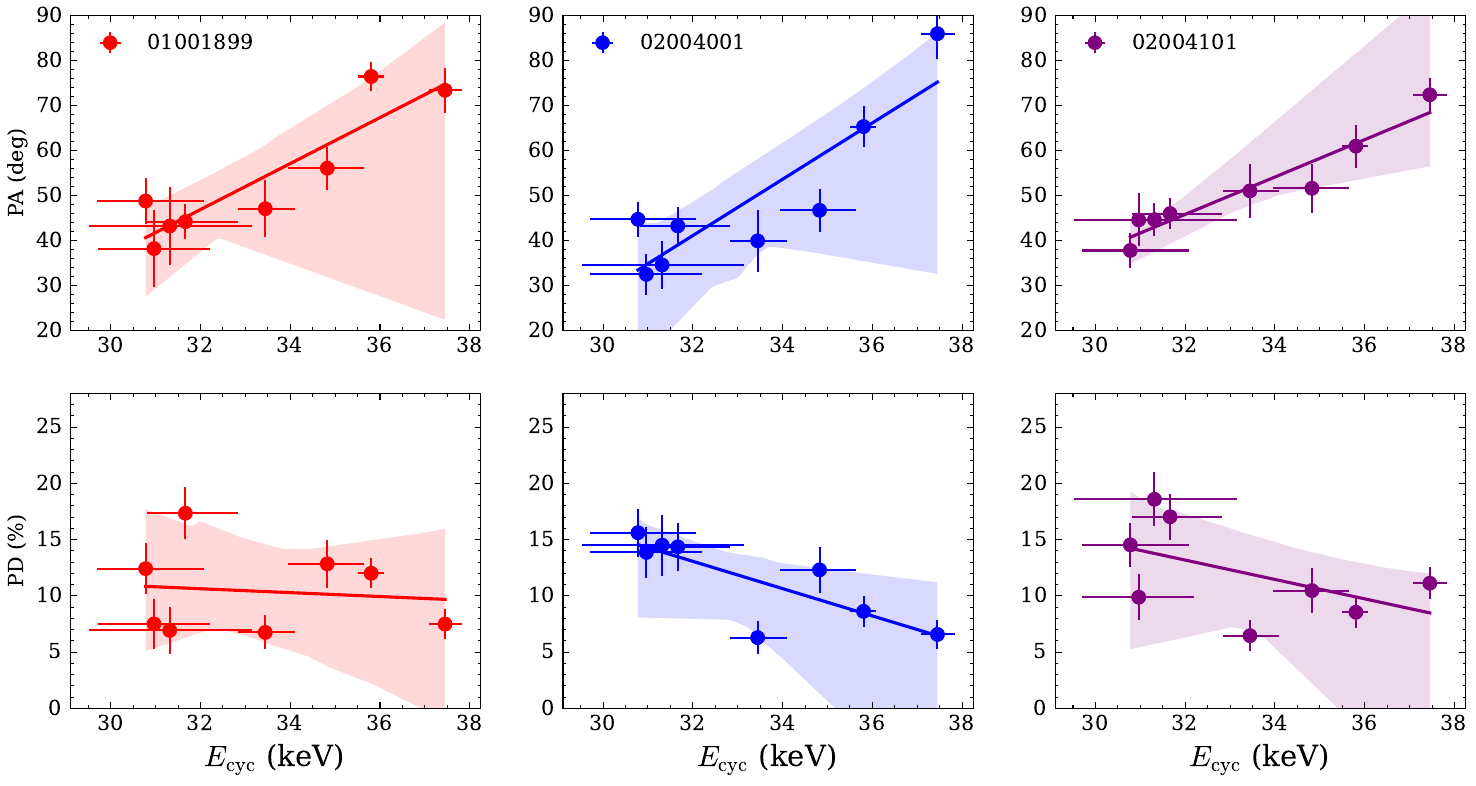}
    \caption{Top panels: the PA versus the magnetic strength $E_{\rm cyc}$. $E_{\rm cyc}$ is from \textit{NuSTAR} ObsID 30601012002. The PA is from Main-on state \textit{IXPE} observations: 01001899, 02004001, 02004101, and is estimated by the \texttt{PCUBE}. The color line and colored shaded regions represent the linear regression and 3$\sigma$ confidence intervals of the fittings. The Pearson correlation coefficients are 0.90, 0.88 and 0.95 for the left, middle and right panels, respectively. The $p$-values are 0.002, 0.004 and 0.0002 for the left, middle and right panels, respectively. Bottom panels: the PD versus $E_{\rm cyc}$. The Pearson correlation coefficients are -0.11, -0.81, -0.51 for the left, middle and right panels. The $p$-values are 0.79, 0.02 and 0.19 for the left, middle and right panels, respectively.}
    \label{fig:fig11}
\end{figure*}

We also divide the Main-on state observations of the two \textit{NuSTAR} datasets into multiple phase bins, corresponding to the \textit{IXPE} Main-on state observations. We note that, in some phases, such as the phase of 0--0.125, the "10\,keV" feature is clearly observed in the spectra. This feature appears as a bump in the 10--20\,keV band in the residuals, as shown in the middle panel of Figure~\ref{fig:fig_10keV}. The physical origin of this feature is not yet well understood \citep{Vasco_spectra}. To model this feature, we added a \texttt{Gabs} model in the spectral fitting, and the residuals are shown in the bottom panel of Figure~\ref{fig:fig_10keV}. The parameters of the cyclotron line are not significantly affected by the inclusion of this component. Due to the relatively low count rate in the Short-on state, we are unable to perform a phase-resolved analysis for this state. In Figure~\ref{fig:fig10}, we present the profiles and variations of the cyclotron line energy as a function of the pulse phase. The cyclotron line energy exhibits a pattern roughly following the pulse profile, with higher values observed around the peak, which is consistent with the results reported by \citet{Vasco_spectra} and \citet{Furst_nustar}. \citet{Vasco_spectra} analyzed multiple \textit{RXTE} observations across various Main-on phases, and their findings demonstrated consistent evolutionary trends with pulse phase for different $\phi_{35}$ phases in one super-orbit cycle. Our results are in good agreement with the studies by \citet{Vasco_spectra} and \citet{Furst_nustar}, suggesting that the variation of the cyclotron line energy does not undergo significant changes over different 35 cycles. Based on these consistent results, it is reasonable to expect that the observed variation during the time span of the \textit{IXPE} observations would exhibit similar behavior.
We also attempt to fit the variations of cyclotron line energy as a function of the pulse phase with the simple dipole model \citep{Suchy_dipole_crsf}. The expression is:
\begin{equation}
B = \frac{B_{0}}{2} \sqrt{1+3\cos{\psi}^2},
\end{equation}
\begin{equation}
\cos{\psi} = \cos{i_{\rm p}}\cos{\theta}+\sin{i_{\rm p}}\sin{\theta}\cos({\phi - \phi_{0}}),
\end{equation}
where $B_{0}$ is the field strength at the magnetic pole, $\psi$ is the angle between the line of sight and the magnetic axis direction, $i_{\rm p}$ represents the angle between the pulsar spin vector and the line-of-sight, $\theta$ is the magnetic obliquity (i.e., the angle between the magnetic dipole and the spin axis), $\phi$ corresponds to the pulse phase, and $\phi_{0}$ indicates the phase when the magnetic pole is closest to the observer. Combining the following equation:

\begin{equation}\label{eq:B12Ecyc}
B_{12} = \frac{E_{\mathrm{cyc}}}{11.6 \, \mathrm{keV}},
\end{equation}
where $B_{12}$ is the magnetic field strength in units of $10^{12}$ G, we can get:

\begin{equation}
E_{\mathrm{cyc}} = \frac{E_{0}}{2} \sqrt{1+3\cos{\psi}^2},
\end{equation}
where $E_{0} = B_{0}\times 11.6/10^{12}$ (keV).
As shown in Figure~\ref{fig:fig10}, the simple dipole model can provide a relatively good fit for the variations of cyclotron line energy over different pulse phases. The fitting is conducted using the \texttt{EMCEE} package, which is an affine invariant Markov Chain Monte Carlo ensemble sampler implemented in Python \citep{Foreman-Mackey_emcee}. Due to the relatively large data errors, to better constrain the parameters, we set prior ranges for the pulsar inclination $i_{\rm p}$ and $\phi_{\rm 0}$ as follows: $[20^{\circ}, 160^{\circ}]$ and [0.8, 1.0], respectively. The posterior distributions are plotted in Figure~\ref{fig:fig14}. The fitting results are summarized in Table.~\ref{table:table3}. The values of $i_{\rm p}$ and $\theta$ are roughly consistent in both of the \textit{NuSTAR} observations, 30601012002 and 90701330002. Therefore, with the simple dipole model, we can loosely constrain the geometry of this system. The parameters $\theta$ and $i_{\rm p}$, as determined by the two \textit{NuSTAR} observations, lie within the ranges of $[8.5^{\circ}, 14.8^{\circ}]$ and $[34^{\circ}, 79^{\circ}]$, respectively.

As illustrated in the top panels of Figure~\ref{fig:fig11}, a distinct positive correlation is evident between PA and cyclotron line energy. The Pearson correlation coefficients are 0.90, 0.88 and 0.95 for the left, middle and right top panels of Figure~\ref{fig:fig11}, respectively. The $p$-values are 0.002, 0.004 and 0.0002 for the left, middle and right top panels of Figure~\ref{fig:fig11}, respectively. Although we only show the correlations between ObsID 30601012002 and the three Main-on \textit{IXPE} observations, the correlations also stand between ObsID 90701330002 and the \textit{IXPE} observations.

\subsection{Fitting with Rotating Vector Model}
\label{sec:results:RVM}
\begin{table*}
\begin{tabular}{l|l|l|l|l|l|l|l|l|}
\hline
\textbf{Instrument} & \textbf{Observation}                   & \textbf{$i_{\rm p}$($^{\circ}$)}         & \textbf{$\theta$ ($^{\circ}$)}   & \textbf{$\chi_{\rm p}$($^{\circ}$)}     & \textbf{$\phi_{0}$/2$\pi$} &  \textbf{$E_{0}$ (keV)} \\ \hline
\multicolumn{1}{|l|}{IXPE} & 01001899 & $68^{+47}_{-35}$ & $14.5^{+3.8}_{-5.5}$ & $54.6^{+2.2}_{-2.2}$ & $0.25^{+0.03}_{-0.03}$ &  -  \\
\multicolumn{1}{|l|}{} & 02003801 &  $90^{+43}_{-43}$ & $2.1^{+2.8}_{-1.5}$  & $42.5^{+2.9}_{-2.9}$ & $0.50^{+0.32}_{-0.30}$ & - \\ 
\multicolumn{1}{|l|}{} & 02004001 &  $45^{+30}_{-22}$ & $12.0^{+4.2}_{-5.3}$ & $49.2^{+1.7}_{-1.7}$ & $0.19^{+0.03}_{-0.02}$ & - \\ 
\multicolumn{1}{|l|}{} & 02003901 &   $90^{+43}_{-43}$ & $2.9^{+3.0}_{-2.0}$ & $41.4^{+2.2}_{-2.1}$ & $0.17^{+0.65}_{-0.11}$ & - \\ 
\multicolumn{1}{|l|}{} & 02004101 &   $109^{+33}_{-38}$ & $12.2^{+3.0}_{-4.5}$ & $51.7^{+1.5}_{-1.5}$ & $0.24^{+0.02}_{-0.02}$ & - \\ \hline
\multicolumn{1}{|l|}{NuSTAR} & 30601012002  &   $74^{+6}_{-25}$ & $10.8^{+3.0}_{-2.1}$ & - &  $0.95^{+0.01}_{-0.01}$ & $59^{+3}_{-15}$ \\
\multicolumn{1}{|l|}{} & 90701330002  &   $68^{+11}_{-34}$ & $10.8^{+4.0}_{-2.3}$ & - &  $0.92^{+0.02}_{-0.02}$ & $57^{+6}_{-17}$ \\
\hline
\end{tabular}
\caption{RVM and simple dipole model fitting results. $i_{\rm p}$ represents the angle between the pulsar spin vector and the line-of-sight. $\theta$ is the magnetic obliquity (i.e., the angle between the magnetic dipole and the spin axis). $ \chi_{\rm p}$ is the position angle of the pulsar spin axis. $\phi_{0}$ indicates the phase when the magnetic pole is closest to the observer. $E_{0}$ is the cyclotron line energy at the magnetic pole.}
\label{table:table3}
\end{table*}

We employ the RVM model to fit the variations of the PA with pulse phase \citep{Radhakrishnan_RVM, Poutanen_RVM, Victor_Herx-1}. In the magnetosphere, the radiation propagates in two modes due to vacuum birefringence \citep{Gnedin_vacuum}. This propagation continues until reaching the polarization-limiting radius, estimated to be around 20 stellar radii (approximately 250 km) for typical X-ray pulsars \citep{Budden_QED, Heyl_QED, Heyl_QED2}. This radius is much larger than the size of the star, implying a dipole field configuration in that region. 

In the RVM, if radiation escapes in the O-mode, PA can be described by the following equation:

\begin{equation}
\tan(\text{PA} - \chi_{\mathrm{p}}) = \frac{-\sin{\theta}\sin(\phi - \phi_{0})}{\sin{i_{\mathrm{p}}}\cos{\theta} - \cos{i_{\mathrm{p}}}\sin{\theta}\cos(\phi - \phi_{0})},
\end{equation}
where $ \chi_{\rm p}$ is the position angle of the pulsar's spin axis.

The RVM model has successfully applied to the phase-dependent PA variation in several X-ray accreting pulsars (Her X--1 \citep{Victor_Herx-1,Heyl_herX-1}, Cen X--3 \citep{Tsygankov_cenX-3}, GRO J1008--57 \citep{Tsygankov_groj1008}, X Persei \citep{Mushtukov_xpeisei}, EXO 2030+375 \citep{Malacaria_exo2030}, GX 301--2 \citep{Suleimanov_gx301}, 1A 0535+262 \citep{Long_0535}, LS V +44 17 \citep{Victor_LSV44}) to determine the pulsar geometry parameters. 

\begin{figure*}
    \centering
    \includegraphics[width=0.55\textwidth]{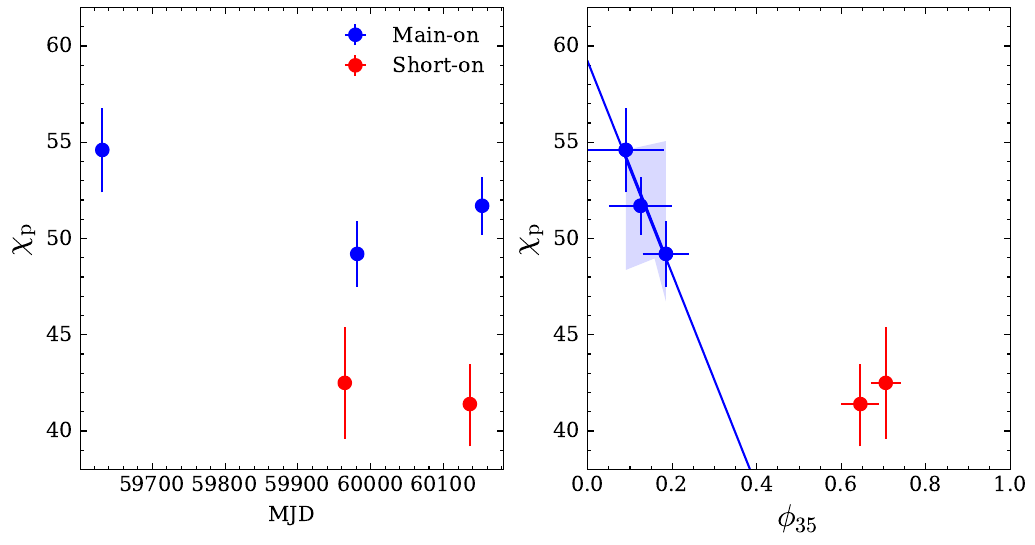}
    \caption{Left panel: The position angle of the spin axis on the sky ($\chi_{\rm p}$) versus MJD. The blue points and red points indicate the Main-on observations and Short-on observations. Right panel: $\chi_{\rm p}$ versus the super-orbital phase ($\phi_{35}$) the slope and intercept are -55 and 59, respectively. The Pearson correlation coefficient and the $p$-value are -0.98 and 0.12, respectively.}
    \label{fig:fig12}
\end{figure*}

We employ the RVM and performed fitting to the pulse phase dependence of the PA for all five \textit{IXPE} observations. The fitting results are summarized in Table~\ref{table:table3}. The posterior distributions for the five observations are plotted in Figures~\ref{fig:fig15}. As shown in Figure~\ref{fig:fig7}, the RVM model appears to provide a relatively good fit for all five \textit{IXPE} observations. For the ObsID 01001899, our results are consistent with the results presented in \citet{Victor_Herx-1} and \citet{Heyl_herX-1} by considering the uncertainties. $\theta$ of the Main-on state observations are larger than the Short-on state observations. $\theta$ can reflect the modulation amplitude of the PA. It is evident that the PA modulations of the Short-on state are weaker compared to the Main-on state, as shown in Figure~\ref{fig:fig7}. The values of $\theta$ and $i_{\rm p}$ of Main-on observations estimated by the RVM fall in the range of $[6.7^{\circ},18.3^{\circ}]$, $[23^{\circ},142^{\circ}]$ respectively, which are roughly consistent with the values determined by the simple dipole model. Although the values of $i_{\rm p}$ display discrepancies across different observations, we caution not to over-interpret this parameter as its uncertainties are very large. As depicted in the left panel of Figure~\ref{fig:fig12}, changes in the position angle of the spin axis on the sky ($\chi_{\rm p}$) across the three Main-on state \textit{IXPE} observations are observed. Shifting from ObsID 01001899 to ObsID 02004001, the $\chi_{\rm p}$ values undergo an approximate change of 5.4$^\circ$. Furthermore, from ObsID 02004001 to ObsID 02004101, this value underwent an alteration of 2.5$^\circ$. In addition, to examine the correlation between $\chi_{\rm p}$ and $\phi_{35}$, we plot the right panel of Figure~\ref{fig:fig12}. For the Main-on observations, a potential linear anti-correlation is observed. We further perform a linear regression, yielding a slope of $-55$ and an intercept of 59. The Pearson correlation coefficient and $p$-value are $-0.98$ and 0.12, respectively. It is worth noting that since there are only three data points in this line, further Main-on observations may be required to thoroughly test this correlation.

\begin{figure*}
    \centering
    \includegraphics[width=0.85\textwidth]{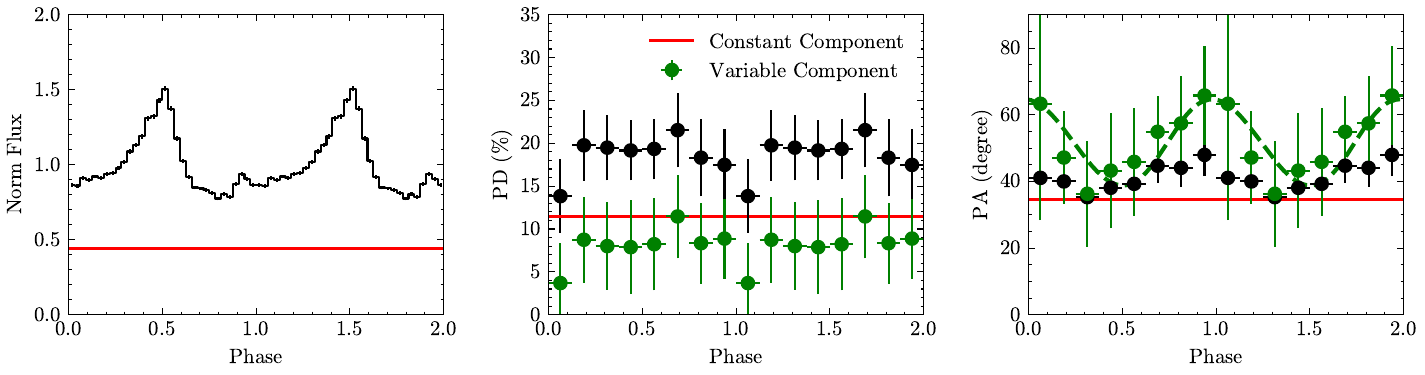}
    \caption{Variations of the pulse profile (left panel), PD (middle panel), and PA (right panel) with pulse phase. The data is from ObsID 02003901 and is shown in black. The red and green symbols represent the constant and variable components. The pulse profile and PD are normalized with the average flux.}
    \label{fig:two_component}
\end{figure*}

\section{DISCUSSION}
\label{sec:discussion}
The detailed polarization analysis of the initial \textit{IXPE} observation for Her X--1 has been presented by \citet{Victor_Herx-1}. The researchers well constrained the geometry parameters using the RVM. \citet{Garg_herx-1} reported a higher PD on the Short-on state compared to the Main-on state. \citet{Heyl_herX-1} revealed the free precession of the neutron star crust through the change in $\theta$ between the Main-on and Short-on states. In this paper, we use all five \textit{IXPE} public data to further explore the polarization properties. 

The unexpected low PD observed in Her X--1 is inconsistent with the theoretical predictions \citep{Caiazzo_polarization_model}. The polarization of intrinsic emission could be significantly modified by the structure of the neutron star's atmosphere. A toy model was proposed by \citet{Victor_Herx-1} to account for the low PD observed in Her X--1. The depolarization could be caused by mode conversion due to vacuum resonance located within an overheated transition layer. Additionally, the observed emission may originate from a combination of emissions from distinct hot spots, and the mixing of these emissions could potentially lead to depolarization. 

A higher PD is observed in the Short-on state compared to those measured in the Main-on state. While the nature of 35-day modulation remains under debate, it is believed to be associated with the precessing warped disk. The neutron star is directly visible in the Main-on state but partially obscured in the Short-on state \citep{Scott_35days}. \citet{Garg_herx-1} suggested that the higher PD is caused by the preferential obstruction of one of the magnetic poles of the pulsar during the Short-on state. Interestingly, the PD and flux, as well as the PD and PF, exhibit anti-correlations, as shown in Figure~\ref{fig:fig6}. In addition, we also notice that a recent work by \citet{Victor_LSV44} on RX J0440.9+4431 emphasized the impact of the polarized component from scattering in the disk wind, which could significantly alter the PA variations with pulse phase. They applied a two-component polarization model and obtained consistent geometry parameters in two observations by introducing a constantly and highly polarized component caused by scattering; whereas without adding such a component, the fitted magnetic obliquity between the two observations would differ approximately $27^{\circ}$. Actually, the structure of the disk wind in Her X--1 has been unveiled by \citet{Kosec_wind} through the investigations of X-ray absorption lines. In view of this, the scattering in the disk wind may potentially impact the observed polarization properties of Her X--1. The polarization of the scattered component relies on the inclination to the plane normal, denoted as "$i$", and can be expressed as ${\rm PD} = \sin^2 i/(3 - \cos^2 i)$ \citep{Sunyaev_wind}. Taking into account the orbital inclination of 85$^{\circ}$ for Her X-1, the PD from the scattered component can reach up to $\sim$ 33\%. The higher PD of the Short-on states and the correlations depicted in Figure~\ref{fig:fig6} could be qualitatively explained by the presence of two polarized components. Specifically, one component originates from the neutron star with higher PF and lower PD, while the other, possibly produced from the scattering in the disk wind, exhibits lower PF and higher PD. The varying contributions of these two components to the overall flux, modulated by the precession of the warped disk, may result in the evolution of both the PF and PD. If the neutron star is less obscured by the warped disk, the total flux would increase, and the proportion of direct emission from the neutron star will also increase, leading to an increase in the PF and a reduction in the PD, and vice versa. However, it is difficult to precisely constrain the contribution of each component. 

We attempt to fit a constant component and a modulating component with ObsIDs 02003901 (Short-on) and 02004101 (Main-on) jointly, assuming the same geometry between the two observations, following the method outlined in \citet{Victor_LSV44}. However, the limited statistics prevents us from constraining those two components well at the same time. Therefore, we reduce the degree of freedom by fixing the geometry parameters derived individually from a single RVM fitting of the ObsID 02004101 and applying it to that of the ObsID 02003901. We then only fit three free parameters, $I_{\rm c}$, $Q_{\rm c}$, and $U_{\rm c}$, for the constant component with ObsID 02003901. The posterior distribution is shown in Figure~\ref{fig:fig16}. In Figure~\ref{fig:two_component}, we plot the PD and PA variations with pulse phase for each component in the middle and right panels. The PA of the constant component is approximately $34.4^{\circ}$. As $I_{\rm c}$ is not well constrained, the uncertainty of the PD of the constant component propagates accordingly. If we take the median value of 0.44 for $I_{\rm c}$, the PD of the constant component is approximately 25.9\%. However, it should be noted that this fitting does not consider the geometry change caused by the free precession between ObsIDs 02003901 and 02004101. The effects caused by the free precession and constant component could be coupling.

We also noted that \citet{Heyl_herX-1} concluded that the scattering does not significantly contribute to the observed emission during the Short-on state. However, an accurate assessment of the impact of the scattering on the observed polarized radiation requires a combination of simultaneous observations in high-resolution spectroscopy and polarimetry, which are currently not available. In this paper, we take a more cautious attitude and refrain from extensive discussion on the RVM fitting parameters obtained under the Short-on states. Additionally, it is important to note that the interpretation of two polarized components to explain the anti-correlations between the flux and PD, as well as the PF and PD, is relatively simplified, overlooking the potential influence of free and forced precession in this system \citep{Heyl_herX-1}.

In addition, the observed polarization is also closely linked to the beam pattern, as discussed by \citet{Meszaros_beam_pattern}. Generally, the pencil beam pattern tends to yield lower polarization values \citep{Meszaros_beam_pattern, Marshall_4U1627}. In the sub-critical state, when the pencil beam predominates, the PD tend likely to decrease as the flux/PF increases during the Main-on observations. \citep{Meszaros_beam_pattern, Tsygankov_cenX-3}. Furthermore, by examining Figure~\ref{fig:fig7}, the PD variations of the Main-on states exhibit a modulation with the pulse phase. Specifically, the PD reaches its minimum when the flux attains its maximum. This is in line with the forecasts of \citet{Meszaros_beam_pattern} for scenarios involving a pencil beam pattern, and is similar to that of Cen X--3 \citep{Tsygankov_cenX-3}.

The RVM model has been employed by \citet{Victor_Herx-1} to derive pulsar geometry parameters on Her X-1. In a related study, \citet{Heyl_herX-1} unveiled the precession of the neutron star crust by employing the RVM model on the initial three observations of Her X--1. Here we apply the RVM model to all five \textit{IXPE} observations (Table~\ref{table:table3} and Figure~\ref{fig:fig7}). The magnetic obliquity $\theta$ exhibits considerable differences between the Main-on and Short-on states, but does not change significantly between different Main-on states. \citet{Heyl_herX-1} suggested that the change of $\theta$ between the Main-on and Short-on states could serve as evidence of free precession of the neutron star crust. As shown in the left panel of Figure~\ref{fig:fig12} and Table~\ref{table:table3}, we observed the changes of $\chi_{\rm p}$ between different Main-on observations. Such changes suggest the presence of forced precession, potentially driven by the combined torque from the warped accretion disk and the torque due to the interaction between the superfluid core and the crust. In addition, as shown in the right panel of Figure~\ref{fig:fig12}, we also observe a linear anti-correlation between $\chi_{\rm p}$ and $\phi_{\rm 35}$ for the Main-on observations, although this correlation is not very significant. Assuming that the forced precession occurs on a 35-day timescale, the required torque would be approximately 5.4 $\times$ $10^{36}$ dyne cm, larger than that provided by the warped disk alone ($\sim$ 2.0 $\times$ $10^{36}$ dyne cm as the upper limit) \citep{Lai_torque,Heyl_herX-1}. One possibility is that this precession may occur on a longer timescale, for example, hundreds of days. \citet{Heyl_herX-1} suggests that this precession may be responsible for the appearance of the anomalous state, which occurs on a five-year timescale \citep{Staubert_profiles}. Additionally, we noticed that $\chi_{\rm p}$ significantly changes ($\sim$ 5 $\sigma$) between ObsIDs 02003901 (Short-on state) and 02004101 (Main-on state). This change of $\chi_{\rm p}$ occurs within approximately 16 days. If the change of $\chi_{\rm p}$ indeed reflects the intrinsic geometry of the neutron star, this would also demand a larger torque beyond the supply from the warped disk alone. As discussed above, the scattering from the disk wind may impact the polarization properties in the Short-on states. This substantial change in $\chi_{\rm p}$ between ObsIDs 02003901 and 02004101 may also reflect the contribution from the scattering in the disk wind.

The $\theta$ and $i_{\rm p}$ values estimated using the cyclotron absorption lines are roughly consistent with those estimated through polarization measurements. Due to non-simultaneous observations and the potential influence of precession effects, it is important to acknowledge that there might be some differences in the parameters estimated using these two methods. Additionally, the simple dipole model has limitations in fitting the cyclotron absorption lines, as it provides little information about how and where the cyclotron line is generated and does not account for influential factors such as gravitational lensing effects. In the case of Her X-1, sinusoidal variations in cyclotron line energy with pulse phase are observed, a pattern that can be approximated by the dipole model. Nevertheless, it is essential to recognize that the magnetic field configuration within the cyclotron line formation region may be considerably more complicated than what the dipole model assumes.  

Despite these limitations, a notable linear correlation between the PA and the cyclotron line energy is observed. The variations in PA can be ascribed to projection effects. Similarly, the variations in cyclotron line energy can be also attributed to alterations in the viewing angle from which the X-ray emitting regions are observed \citep{Kreykenbohm_pulse_phase_crsf}. As a result, the observed correlation between the PA and cyclotron line energy suggests that both phenomena result from viewing angle effects. This correlation reinforces that the pulsar is primarily governed by its dipole magnetic field.

\section{CONCLUSIONS}
\label{sec:conclusion}
In this paper, we present a detailed analysis of Her X--1 with \textit{IXPE} and \textit{NuSTAR}. We identified the anti-correlations between the flux and PD as well as the PF and PD. Besides, we also try to constrain the geometry with cyclotron line and polarization. The geometry parameters $\theta$ and $i_{\rm p}$ estimated by the cyclotron line and polarization are roughly consistent. We find a linear correlation between PA and magnetic field, suggesting the variations of PA and cyclotron line are both viewing angle effects. Additionally, the change of $\chi_{\rm p}$ between different Main-on states suggests the possible forced precession of neutron star crust. To fully understand the intrinsic polarization properties, it is essential to undertake a combined analysis of spectroscopy, timing, and polarimetry. Particularly, improving the statistical significance in polarimetry is of utmost importance. This will be accomplished through upcoming missions like the enhanced X-ray Timing and Polarimetry (\textit{eXTP}) observatory \citep{Zhang_eXTP}.

\section*{Acknowledgements}

We are grateful to the anonymous referee for the valuable comments that have helped improve the paper. Q. C. Zhao and H. C. Li would like to express their sincere gratitude to Victor Doroshenko for his invaluable assistance in replicating the results of ObsID 01001899. This work is supported by the National Key R\&D Program of China (2021YFA0718500). We acknowledge funding support from the National Natural Science Foundation of China (NSFC) under grant Nos. 12122306, 12333007 and 12027803, the CAS Pioneer Hundred Talent Program Y8291130K2 and the Scientific and technological innovation project of IHEP Y7515570U1. H. Feng acknowledges funding support from the National Natural Science Foundation of China under grant No. 12025301. H. C. Li acknowledge the support of the Swiss National Science Foundation.

\section*{Data Availability}

The data utilized in this analysis are publicly accessible via the High Energy Astrophysics Science Archive Research Centre (HEASARC) at \url{https://heasarc.gsfc.nasa.gov/cgi-bin/W3Browse/w3browse.pl}.




\bibliographystyle{mnras}
\bibliography{mnras_template_ref} 





\appendix

\section{supplementary material}
\begin{figure}
    \centering
    \includegraphics[width=0.45\textwidth]{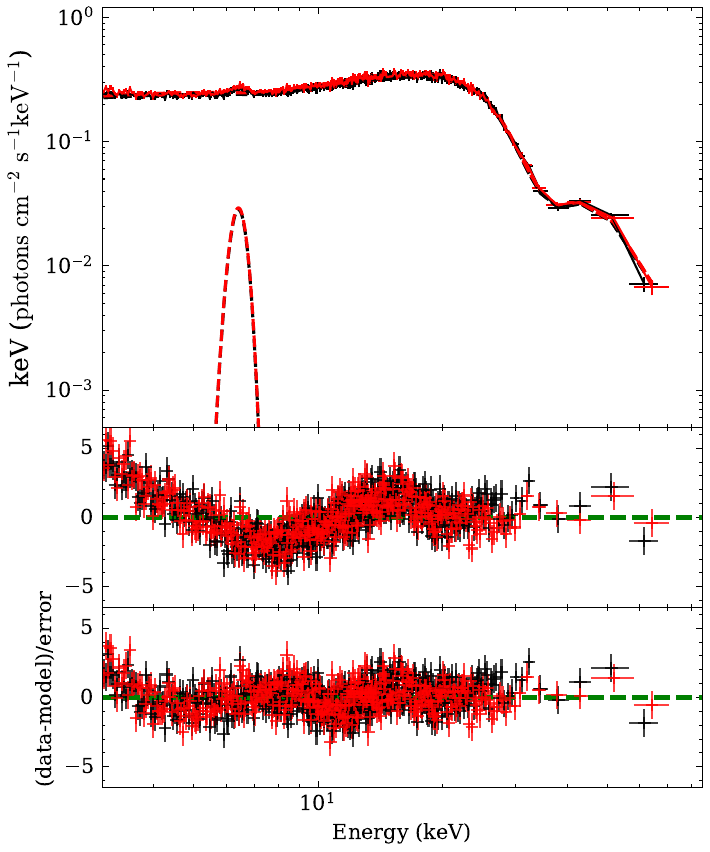}
    \caption{The spectral fitting and residuals at the pulse phase of 0--0.125. The spectral fitting residuals in the middle panel show a bump feature in the 10--20\,keV band. We add a \texttt{Gabs} model to fit this feature, and the fitting residuals are shown in the bottom panel.}
    \label{fig:fig_10keV}
\end{figure}

\begin{figure}
    \centering
    \includegraphics[width=0.4\textwidth]{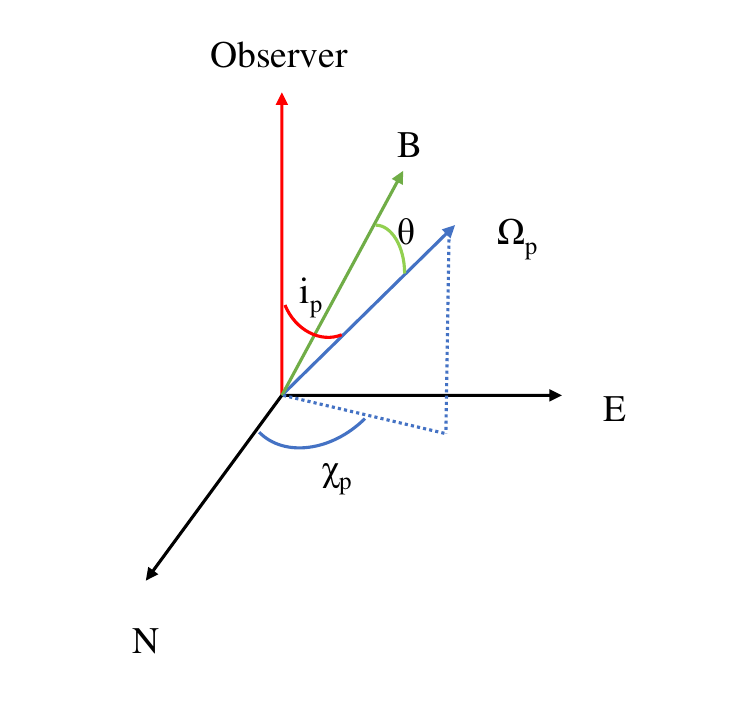}
    \caption{Geometry of the system. Some angles used in the RVM fitting are shown. $i_{\rm p}$ is the angle between the line of sight and the spin axis of the pulsar ($\Omega_{\rm p}$). $\theta$ is the angle between the magnetic dipole (B) and $\Omega_{\rm p}$. $\chi_{\rm p}$ is the position angle defined from the North pole eastward to the projection of $\Omega_{\rm p}$ onto the North-East plane of the sky.}
    \label{fig:fig13}
\end{figure}

\begin{figure*}
    \centering
    \includegraphics[width=0.45\textwidth]{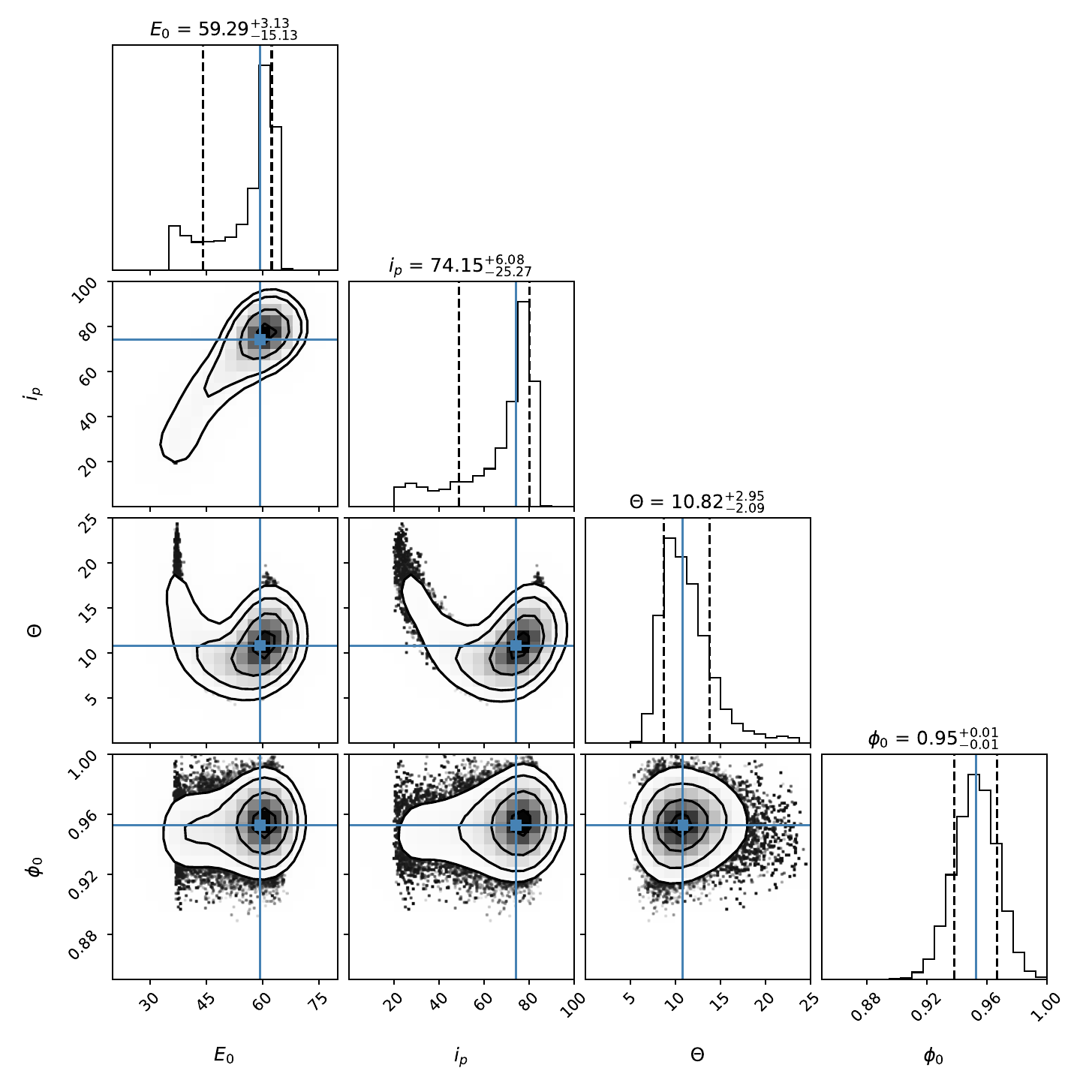}
    \includegraphics[width=0.45\textwidth]{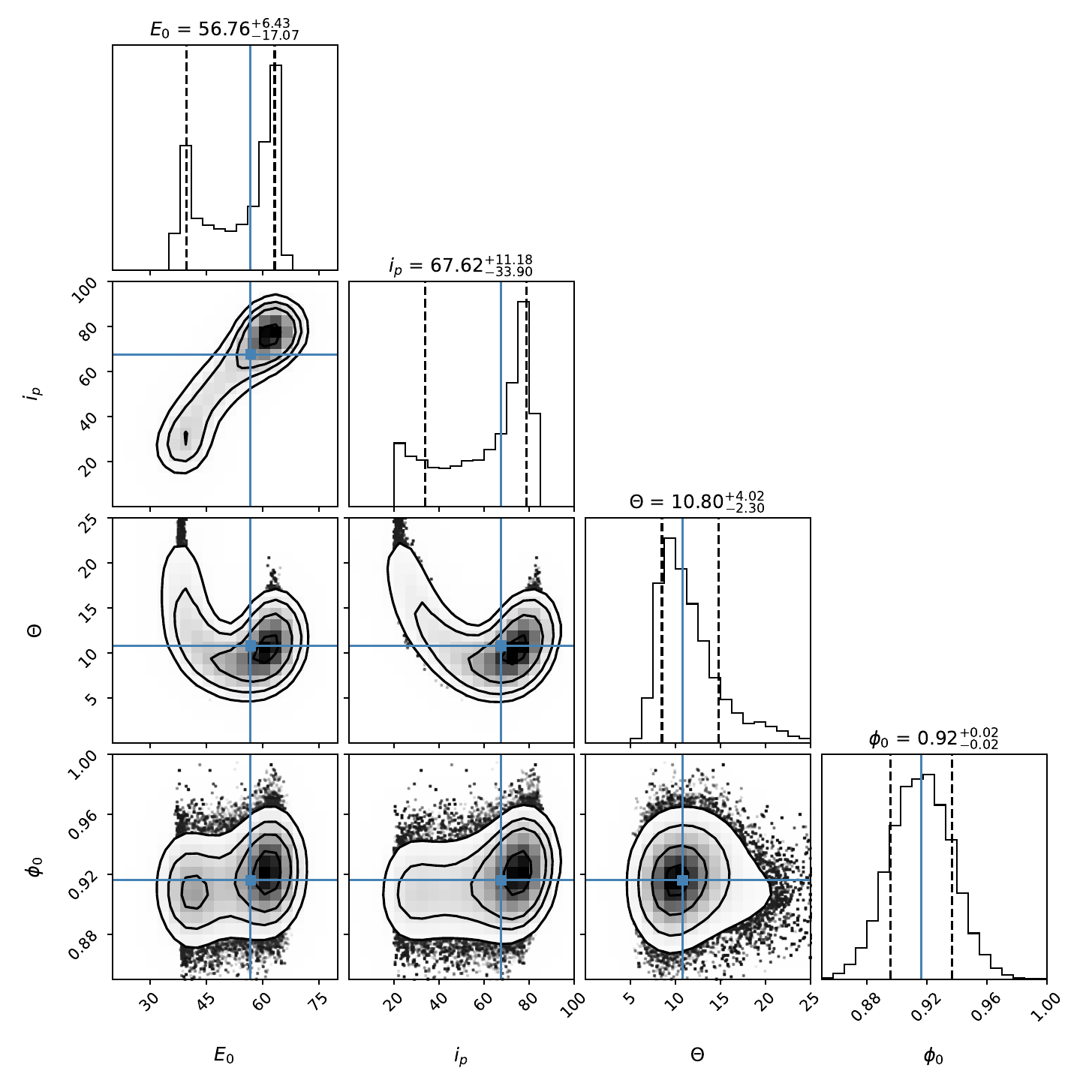}
    \caption{Corner plots of the posterior distribution for the simple dipole model parameters of the \textit{NuSTAR} ObsIDs 306010112002 and 90701330002, respectively.}
    \label{fig:fig14}
\end{figure*}

\begin{figure*}
    \centering
    \vspace{-4mm}
    \includegraphics[width=0.45\textwidth]{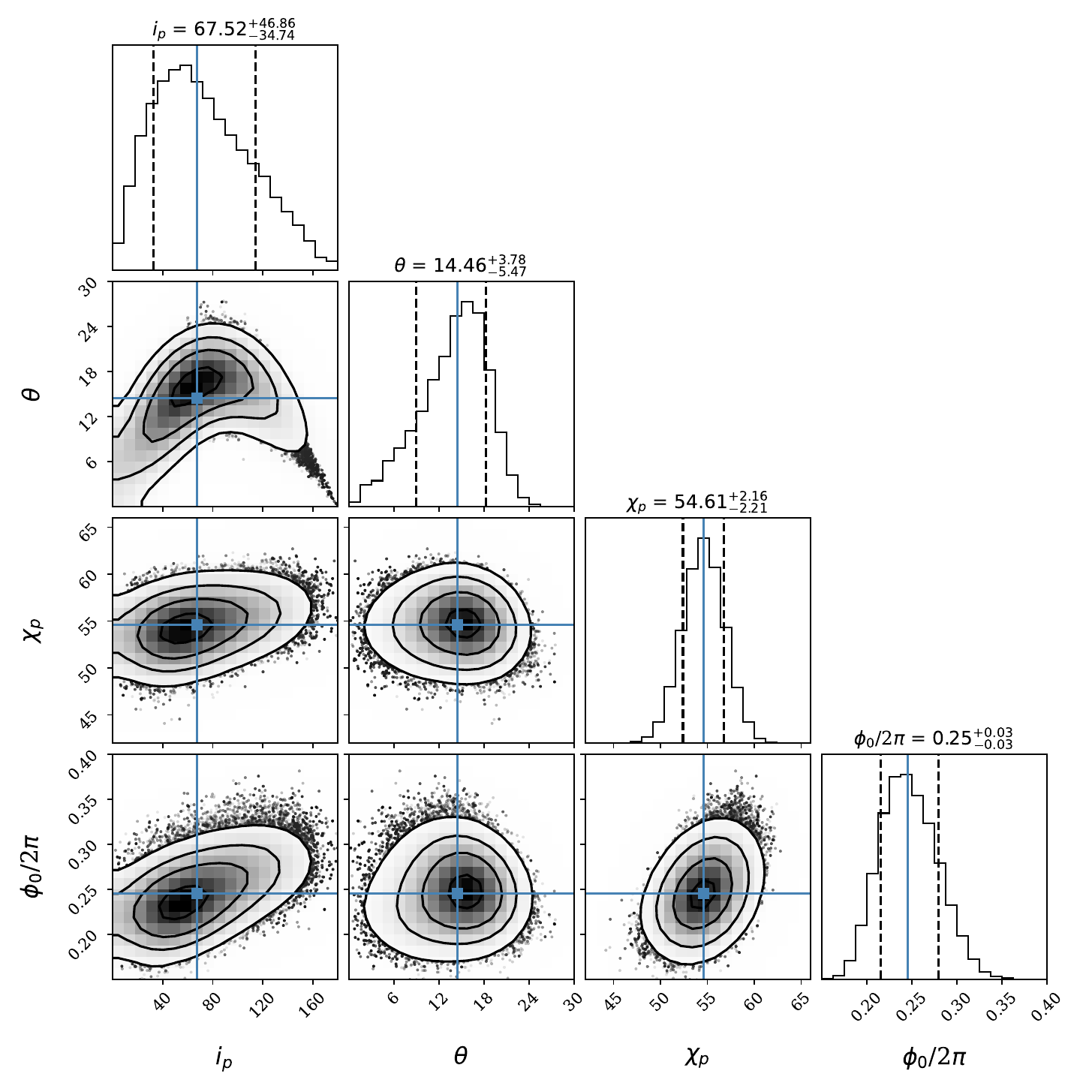}
     \includegraphics[width=0.45\textwidth]{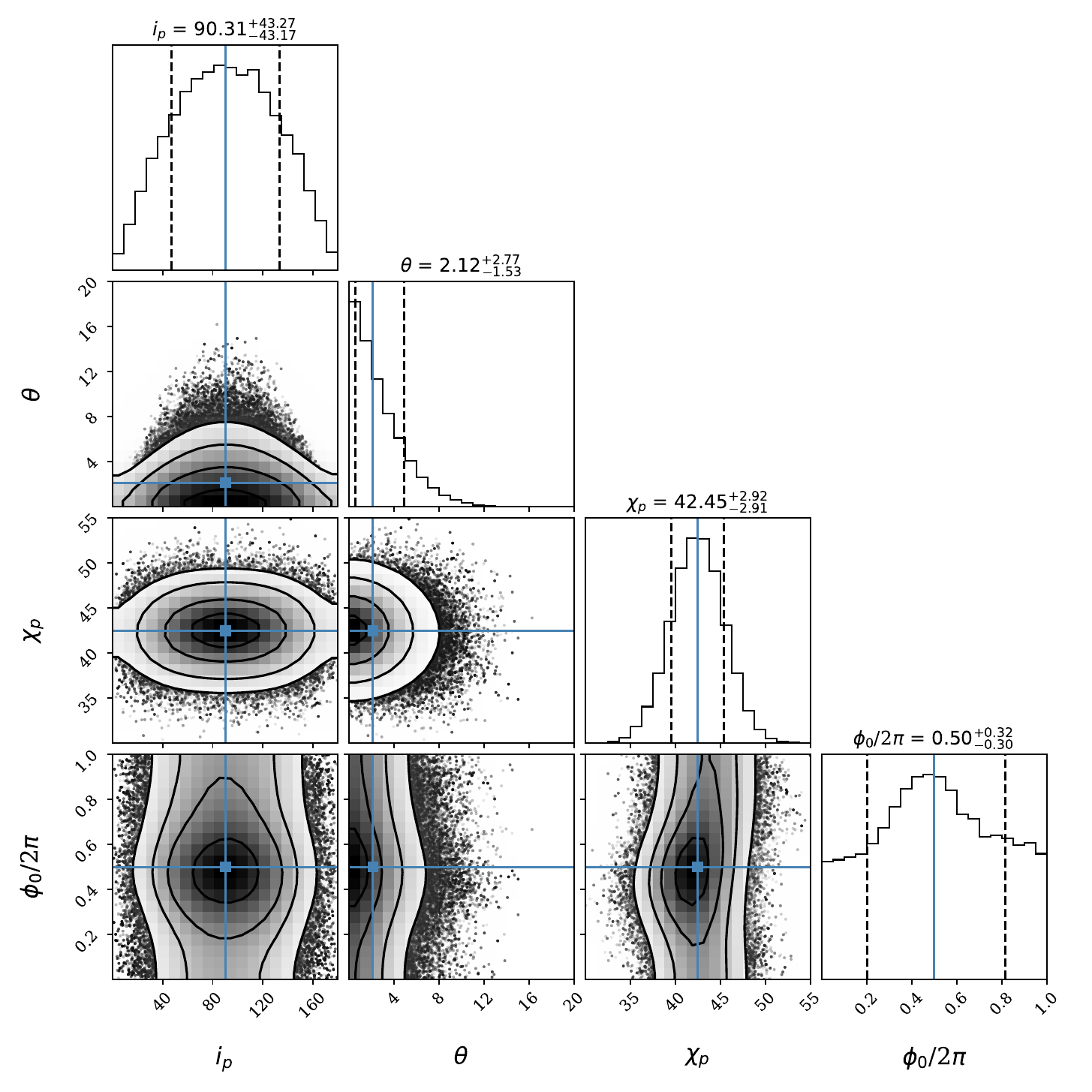}
     \includegraphics[width=0.45\textwidth]{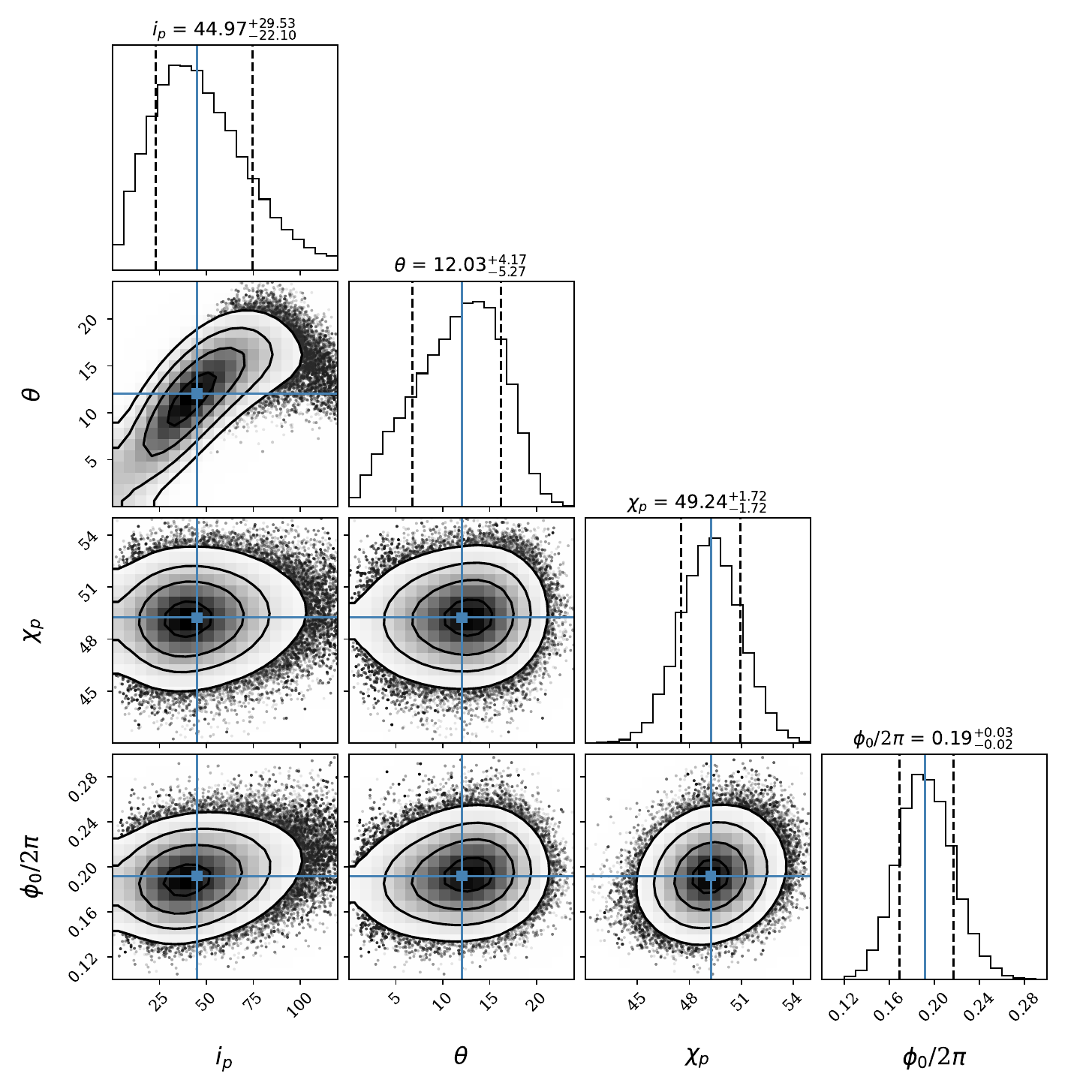}
     \includegraphics[width=0.45\textwidth]{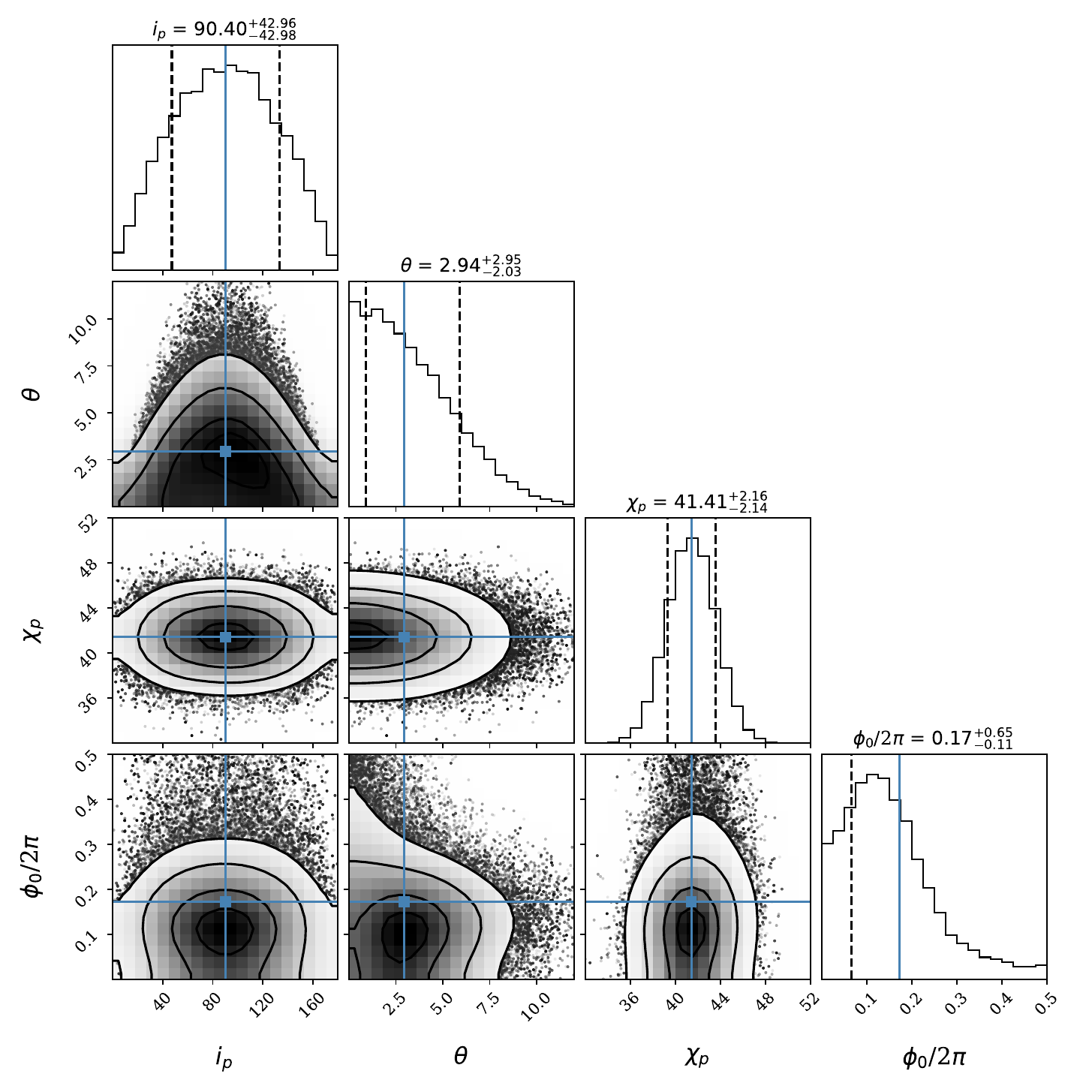}
     \includegraphics[width=0.45\textwidth]{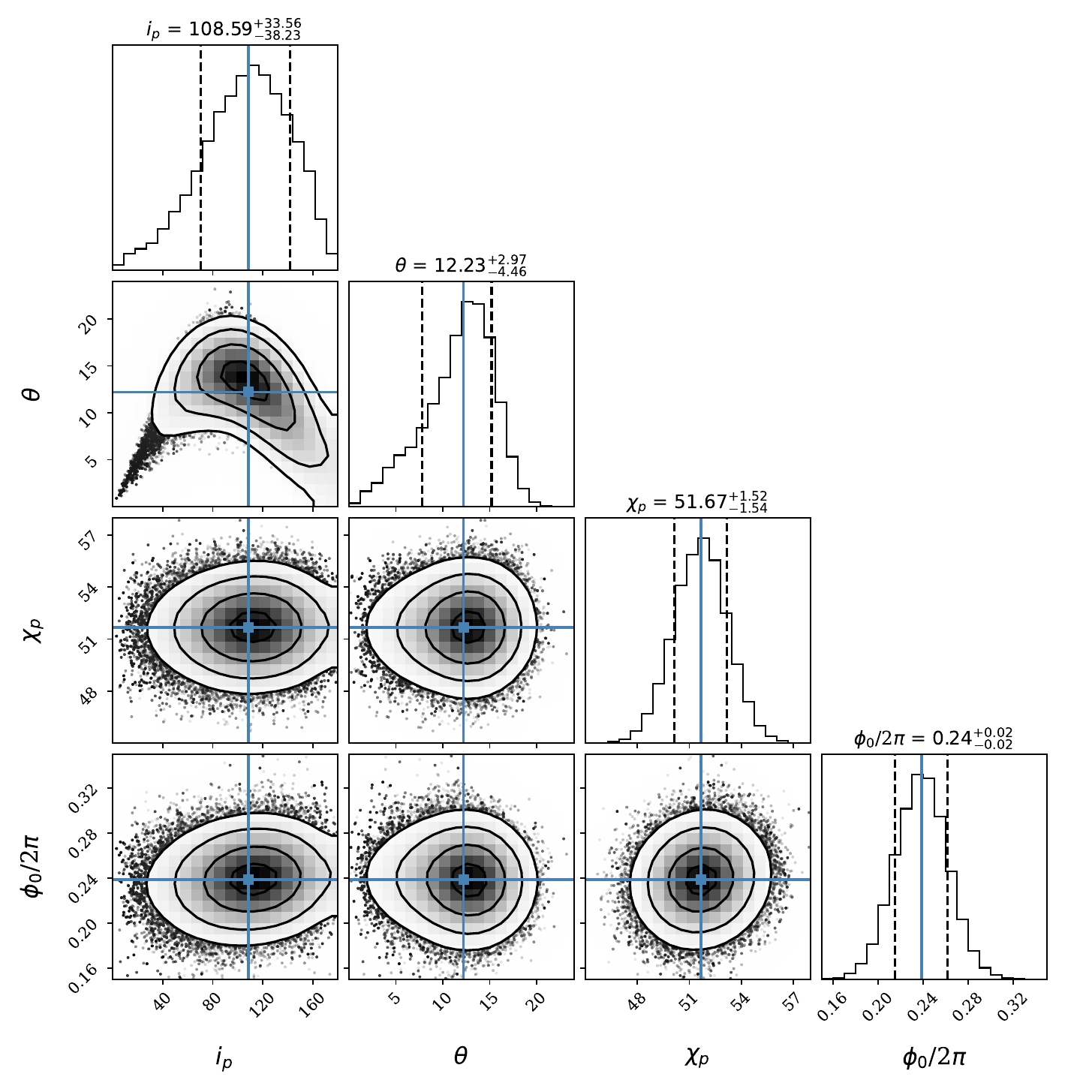}
     \vspace{-2mm}
    \caption{Corner plot of the posterior distribution for the RVM parameters of the IXPE ObsID 01001899, 02003801, 02004001, 02003901 and 02004101.}
    \label{fig:fig15}
\end{figure*}

\begin{figure}
    \centering
    \includegraphics[width=0.45\textwidth]{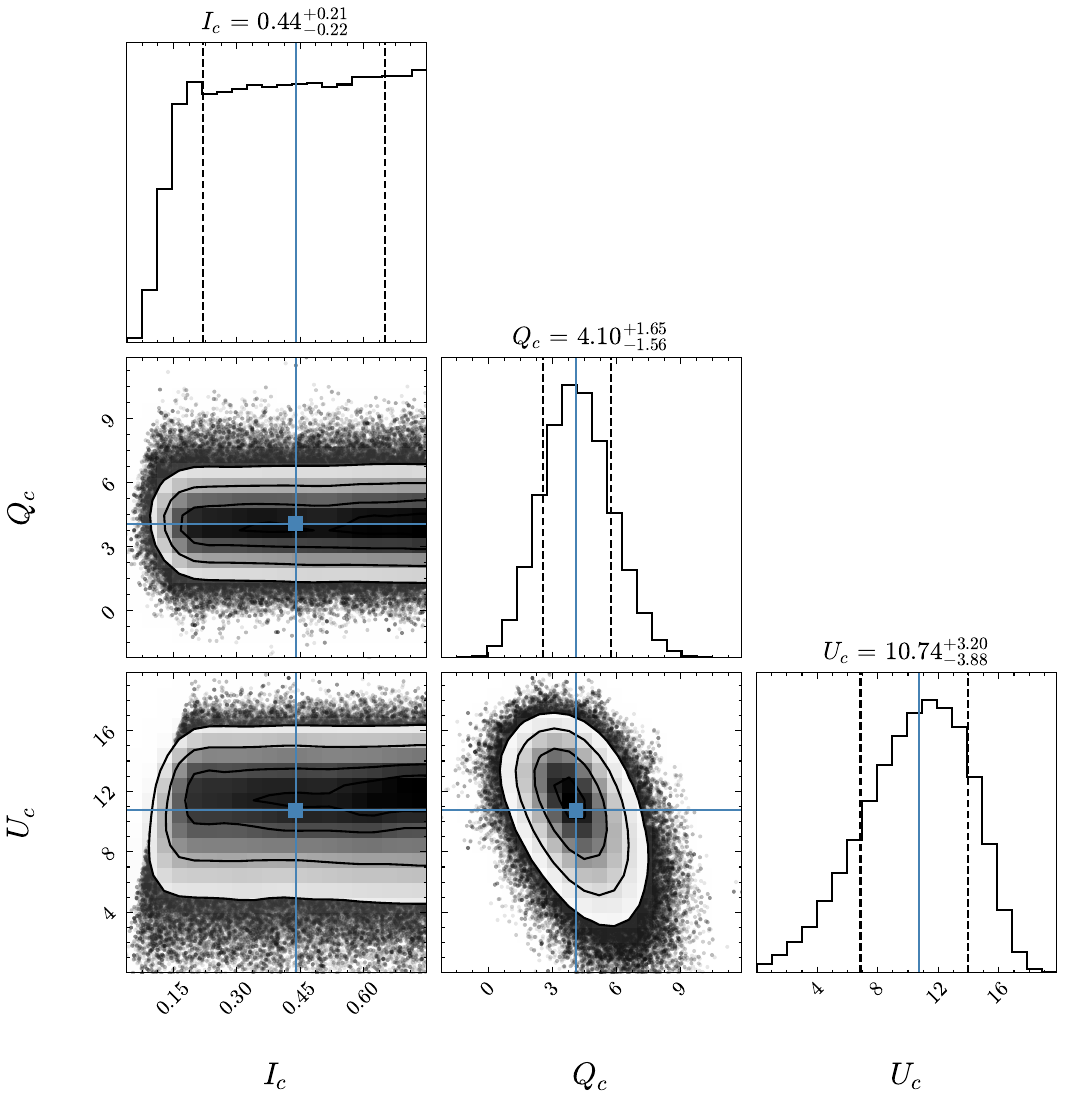}
    \caption{Corner plot of the posterior distribution for the constant component $I_{\rm c}$, $Q_{\rm c}$, and $U_{\rm c}$, which are normalized with the average flux. $Q_{\rm c}$ and $U_{\rm c}$ are multiplied by 100 for clarity.}
    \label{fig:fig16}
\end{figure}

\bsp	
\label{lastpage}
\end{document}